\documentclass[journal]{IEEEtran}

\usepackage{cite} 

\ifCLASSINFOpdf
  \usepackage[pdftex]{graphicx}
  \graphicspath{{images/}}
  \DeclareGraphicsExtensions{.pdf,.jpeg,.png}
\else
  \usepackage[dvips]{graphicx}
  \graphicspath{{images/}}
  \DeclareGraphicsExtensions{.eps}
\fi  
  
\usepackage[cmex10]{amsmath}
\usepackage{amssymb}
\usepackage{amsthm}

\usepackage{algpseudocode} 

\usepackage{array}

\usepackage{mathrsfs}
\DeclareMathAlphabet{\mathpzc}{OT1}{pzc}{m}{it}

\usepackage[caption=false,font=footnotesize]{subfig}

\usepackage{url}

\usepackage{fancybox}

\usepackage{calc}
\newsavebox\myboxa

\usepackage{overpic}
\usepackage{psfrag}

\usepackage{dblfloatfix}

\usepackage{txfonts}
\usepackage{textcomp}

\usepackage{stmaryrd}

\usepackage{pbox}

\usepackage{diagbox}

\usepackage{xcolor}
\usepackage{colortbl}

\usepackage{hhline}

\usepackage{multirow}

\usepackage{makecell}

\makeatletter

\newcommand{\Rmnum}[1]{\expandafter\@slowromancap\romannumeral #1@}
\makeatother

\begin{document}

\title{ Large-scale {JPEG} image steganalysis using hybrid  
  deep-learning framework}

\author{ Jishen Zeng,~\IEEEmembership{Student Member,~IEEE,}
  Shunquan~Tan*,~\IEEEmembership{Senior Member,~IEEE,}
  Bin~Li,~\IEEEmembership{Senior Member,~IEEE,}
  and~Jiwu~Huang,~\IEEEmembership{Fellow,~IEEE}%

  \thanks{This work was supported in part by the NSFC~(61772349,
    U1636202, 61402295, 61572329, 61702340), Guangdong
    NSF~(2014A030313557), Shenzhen R\&D
    Program~(JCYJ20160328144421330). This work was also supported by
    Alibaba Group through Alibaba Innovative Research (AIR) Program.
    \textit{(Corresponding author: Shunquan Tan.)}}%

        \thanks{S.~Tan is with College of Computer Science and
          Software Engineering, Shenzhen University. J.~Zeng, B.~Li,
          and J.~Huang are with College of Information Engineering,
          Shenzhen University.}%

        \thanks{All the members are with Shenzhen Key Laboratory of
          Media Security, Guangdong Province, 518060 China.~(e-mail:
          tansq@szu.edu.cn).}

} 

\maketitle
\begin{abstract}
  Adoption of deep learning in image steganalysis is still in its
  initial stage. In this paper we propose a generic hybrid
  deep-learning framework for JPEG steganalysis incorporating the
  domain knowledge behind rich steganalytic models. Our proposed
  framework involves two main stages. The first stage is hand-crafted,
  corresponding to the convolution phase and the quantization \&
  truncation phase of the rich models. The second stage is a compound
  deep neural network containing multiple deep subnets in which the
  model parameters are learned in the training procedure. We provided
  experimental evidences and theoretical reflections to argue that the
  introduction of threshold quantizers, though disable the
  gradient-descent-based learning of the bottom convolution phase, is
  indeed cost-effective. We have conducted extensive experiments on a
  large-scale dataset extracted from ImageNet. The primary dataset
  used in our experiments contains 500,000 cover images, while our
  largest dataset contains five million cover images. Our experiments
  show that the integration of quantization and truncation into
  deep-learning steganalyzers do boost the detection performance by a
  clear margin. Furthermore, we demonstrate that our framework is
  insensitive to JPEG blocking artifact alterations, and the learned
  model can be easily transferred to a different attacking target and
  even a different dataset. These properties are of critical
  importance in practical applications.
\end{abstract}


\section{Introduction}
\label{sec:intro}

\IEEEPARstart{I}{mage} steganography can be divided into two main
categories: spatial-domain and frequency-domain steganography. The
latter focuses primarily on JPEG images due to their ubiquitous
nature. Both categories in state-of-the-art algorithms adopt
content-adaptive embedding schemes~\cite{fridrich_spie_2007}. Most of
these schemes use an additive distortion function defined as the sum
of embedding costs of all changed elements. From early
HUGO~\cite{pevny_ih_2010}, to latest HILL~\cite{li_icip_2014} and
MiPOD~\cite{sedighi_tifs_2016}, the past few years witnessed the
flourish of additive schemes in spatial domain. In JPEG domain,
UED~\cite{guo_tifs_2014} and UERD~\cite{guo_tifs_2015} are two
additive schemes with good security performance. UNIWARD proposed in
\cite{holub_eurasip_2014} is an additive distortion function which can
be applied for embedding both in spatial and JPEG domains. Its JPEG
version, J-UNIWARD, achieves best
performance~\cite{holub_eurasip_2014,guo_tifs_2015}.  Research on
non-additive distortion functions has made great progress in the
spatial domain~\cite{denemark_ihmmsec_2015,li_tifs_2015}. However,
analogous schemes have not yet been proposed in the JPEG
domain. Although utilizing side information of a pre-cover image~(raw
or uncompressed) can improve the security of JPEG
steganography~\cite{holub_eurasip_2014,guo_tifs_2015,denemark_wifs_2015},
its applicability remains limited due to scarce availability of
pre-cover images.

Most of modern universal steganalytic detectors use a rich model with
tens of thousands of features~\cite{fridrich_tifs_2012,
  denemark_wifs_2014,tang_tifs_2016} and an ensemble
classifier~\cite{kodovsky_tifs_2012}. In spatial domain,
SRM~\cite{fridrich_tifs_2012} and its selection-channel-aware
variants~\cite{denemark_wifs_2014,tang_tifs_2016} reign supreme. In
JPEG domain, DCTR~\cite{holub_tifs_2014} feature set combines
relatively low dimensionality and competitive performance, while
PHARM~\cite{holub_spie_2015} and GFR~\cite{song_ihmmsec_2015} exhibit
better performance, although at the cost of higher dimensionality
w.r.t. DCTR. SCA proposed in \cite{denemark_tifs_2016} is a
selection-channel-aware variant of JPEG rich models targeted at
content-adaptive JPEG steganography~\footnote{Throughout this paper,
  the acronyms used for the steganographic and steganalytic algorithms
  are taken from the original papers. The corresponding full names are
  omitted for brevity.}.

In recent years, with help of parallel computing accelerated by
GPU~(Graphics Processing Unit) and huge amounts of training data, deep
learning frameworks have achieved overwhelming superiority over
conventional approaches in many pattern recognition and machine
learning problems~\cite{schmidhuber_nn_2015}. Researchers in image
steganalysis have also tried to investigate the potential of deep
learning frameworks in this field. Tan et al. explored the application
of stacked convolutional auto-encoders, a specific form of deep
learning frameworks in image steganalysis~\cite{tan_apsipa_2014}. Qian
et al. proposed a steganalyzer based on CNN~(Convolutional Neural
Network) which achieving performance close to
SRM~\cite{qian_spie_2015}, and demonstrated its transfer
ability~\cite{qian_icip_2016}. In \cite{pibre_ei_2016_ver2}, Pibre et
al. revealed CNN based steganalyzers can achieve superior performance
in the scenario that embedding key is reused for different stego
images. Xu et al. constructed another CNN-based
steganalyzer~\cite{xu_spl_2016,xu_ihmmsec_2016} equipped with
BN~(Batch Normalization) layers~\cite{ioffe_arxiv_2015}. Its
performance slightly surpass SRM. In this paper the model proposed by
Xu et al. in \cite{xu_spl_2016} is referred as Xu's model and is used
for detection performance comparison. In \cite{sedighi_ei_2017},
Sedighi and Fridrich implemented a specific CNN layer to imitate rich
steganalytic model but failed to reached state-of-the-art performance.
However, all of the above
approaches~\cite{tan_apsipa_2014,qian_spie_2015,qian_icip_2016,
  pibre_ei_2016_ver2,xu_spl_2016,xu_ihmmsec_2016,sedighi_ei_2017},
focusing on spatial-domain steganalysis, are all evaluated on the
BOSSBase~(v1.01) dataset~\cite{bas_ih_2011}. BOSSBase is arguably not
representative of real-world steganalysis
performance~\cite{sedighi_spie_2016}. With only 10,000 images, deep
learning frameworks trained on BOSSBase are prone to overfitting.
Furthermore, except our work which study the effect of fitting
deep-learning steganalytic framework to a JPEG rich-model features
extraction procedure~\cite{zeng_ei_2017}, no prior works addressed the
application of deep learning frameworks in JPEG steganalysis.

In this paper, we proposed a generic hybrid deep-learning framework
for large-scale JPEG steganalysis. Our proposed framework combines the
bottom hand-crafted convolutional kernels and threshold quantizers
pairing with the upper compact deep-learning model. Experimental
evidences and theoretical reflections are provided to show the
rationale of our proposed framework.  Furthermore, we have conducted
extensive experiments on a large-scale dataset extracted from
ImageNet~\cite{imagenet_website} to demonstrate the capacity of our
proposed generic framework under different scenarios.

The rest of the paper is organized as follows. In
Sect.~\ref{sec:proposed}, we describe the proposed hybrid
deep-learning framework in detail, and provide experimental and
theoretical testimonies to support its rationale. Results of
experiments conducted on large-scale datasets are presented in
Sect.~\ref{sec:exp}. Finally, we make a conclusion in
Sect.~\ref{sec:concluding}.

\section{Our proposed {JPEG} steganalytic framework}
\label{sec:proposed}

In this section, we firstly introduce the training procedure of CNN as
preliminaries. Then we discuss the motivations and challenges related
to the introduction of quantization and truncation in JPEG
deep-learning steganalysis. Finally we describe our generic framework
with experimental evidences and theoretical reflection to support our
design.

\subsection{Preliminaries}
\label{sec:proposed_pre}

The principal part of CNN is a cascade of alternating convolutional
layers, regulation layers~(e.g. BN layers~\cite{ioffe_arxiv_2015}) and
pooling layers. On top of the principal part, there are usually
multiple fully-connected layers. Please note that in CNN, only
convolutional layers and fully-connected layers contain neuron units
with learnable weights and biases~\footnote{The learnable parameters
  \{$\gamma$, $\beta$\} for BN layers are omitted for
  brevity.}. Whether belongs to a convolutional layer or a
fully-connected layer, each neuron unit receives inputs from a
previous layer, performs a dot product with weights and optionally
follows it with a nonlinear point-wise activation function. CNNs can
be trained using backpropagation. For clarity, we omit those layers
without learnable weights and biases, and denote the cascade of layers
with learnable weights and biases in a given CNN as
$[L_1, L_2, \cdots, L_n]$, where $L_1$ is the input layer and $L_n$ is
the output layer. $L_2, \cdots, L_{n-1}$ are the layers whose weights
and biases are trained in backpropagation, namely convolutional layers
and fully-connected layers. Let $a_i^{(l)}$ denote the
activation~(output) of unit $i$ in layer $L_l$. For $L_1$, $a_i^{(1)}$
is the $i$-th input fed to the framework. $W_{ij}^{(l)}$ denotes the
weight associated with unit $i$ in $L_l$ and unit $j$ in $L_{l+1}$,
while $b_j^{(l)}$ denotes the bias associated with unit $j$ in
$L_{l+1}$.  The weighted sum of inputs to unit $j$ in $L_{l+1}$ is
defined as:
\begin{equation}
  \label{eq:2.1-3}
  z_j^{(l+1)}=\sum_{i}W_{ij}^{(l)}a_{i}^{(l)}+b_j^{(l)}
\end{equation}
and $a_j^{(l+1)}=f(z_j^{(l+1)})$ where $f(\cdot)$ is the activation
function.  The set of all $W_{ij}^{(l)}$ and $b_j^{(l)}$ constitutes
the parameterization of a neural network and is denoted as $W$ and
$b$, respectively. For a mini-batch of training features-label pairs
$\{(x^{(1)},y^{(1)}), \cdots, (x^{(m)},y^{(m)})\}$, the goal of
backpropagation is to minimize the overall cost function $J(W,b)$ with
respect to $W$ and $b$:
\begin{equation}
  \label{eq:2.1-4}
  J(W,b)=\frac{1}{m}\sum_{h=1}^m J(W,b;x^{(h)},y^{(h)})+R(W)
\end{equation}
where $R(W)$ is a regularization term which suppresses the magnitude
of the weights, and $J(W,b;x^{(h)},y^{(h)})$ is an error metric with
respect to a single example $(x^{(h)},y^{(h)})$.\footnote{There are
  various forms of $J(W,b;x^{(h)},y^{(h)})$ and $R(W)$, and their
  definitions are omitted here, since irrelevant to the subject of
  this paper.}  For each training sample, the backpropagation
algorithm firstly performs a feedforward pass and computes the
activations for layers $L_2$, $L_3$ and so on up to the output layer
$L_n$. For the $j$-th output unit in the output layer $L_n$, set the
corresponding partial derivative of $J(W,b;x^{(h)},y^{(h)})$ with
respect to $z_j^{(n)}$:
\begin{equation}
  \label{eq:2.1-vartheta_n}
  \vartheta_j^{(n)}=\frac{\partial}{\partial
            a_j^{(n)}}J(W,b;x^{(h)},y^{(h)})f'(z_j^{(n)})
\end{equation}
Then in the backpropagation pass, partial derivatives are propagated
from $L_n$ back to the second last layer $L_2$. For the $j$-th neuron
unit in layer $L_l$, set:
\begin{equation}
  \label{eq:2.1-vartheta_j}
  \vartheta_j^{(l)}=(\sum_{k}W_{jk}^{(l)}\vartheta_{k}^{(l+1)})f'(z_j^{(l)})
\end{equation}
The partial derivatives with respect to $W_{ij}^{(l)}$ and
$b_j^{(l)}$, $l=n-1, n-2, \cdots, 1$ are calculated as:
\begin{equation}
  \label{eq:2.1-5}
  \left\{ \begin{array}{l}
           \frac{\partial}{\partial
           W_{ij}^{(l)}}J(W,b;x^{(h)},y^{(h)})=a_i^{(l)}\vartheta_j^{(l+1)}\
            ,\\
           \frac{\partial}{\partial
            b_j^{(l)}}J(W,b;x^{(h)},y^{(h)})=\vartheta_j^{(l+1)}\ ,
         \end{array}\right.
\end{equation}
Gradient descent is used to find the optimal $W$ and $b$. In the
optimization procedure, it updates $W$ and $b$ according to steps
proportional to the negative of the average of $m$ gradients each of
which is the vector whose components are the partial derivatives in
\eqref{eq:2.1-5}~\cite{cs231n_website}.

\subsection{The introduction of quantization and truncation in
  deep-learning based steganalysis}
\label{sec:proposed_importance_of_qt}

\begin{figure*}[!t]
  \centering
  \includegraphics[width=0.75\linewidth,keepaspectratio]{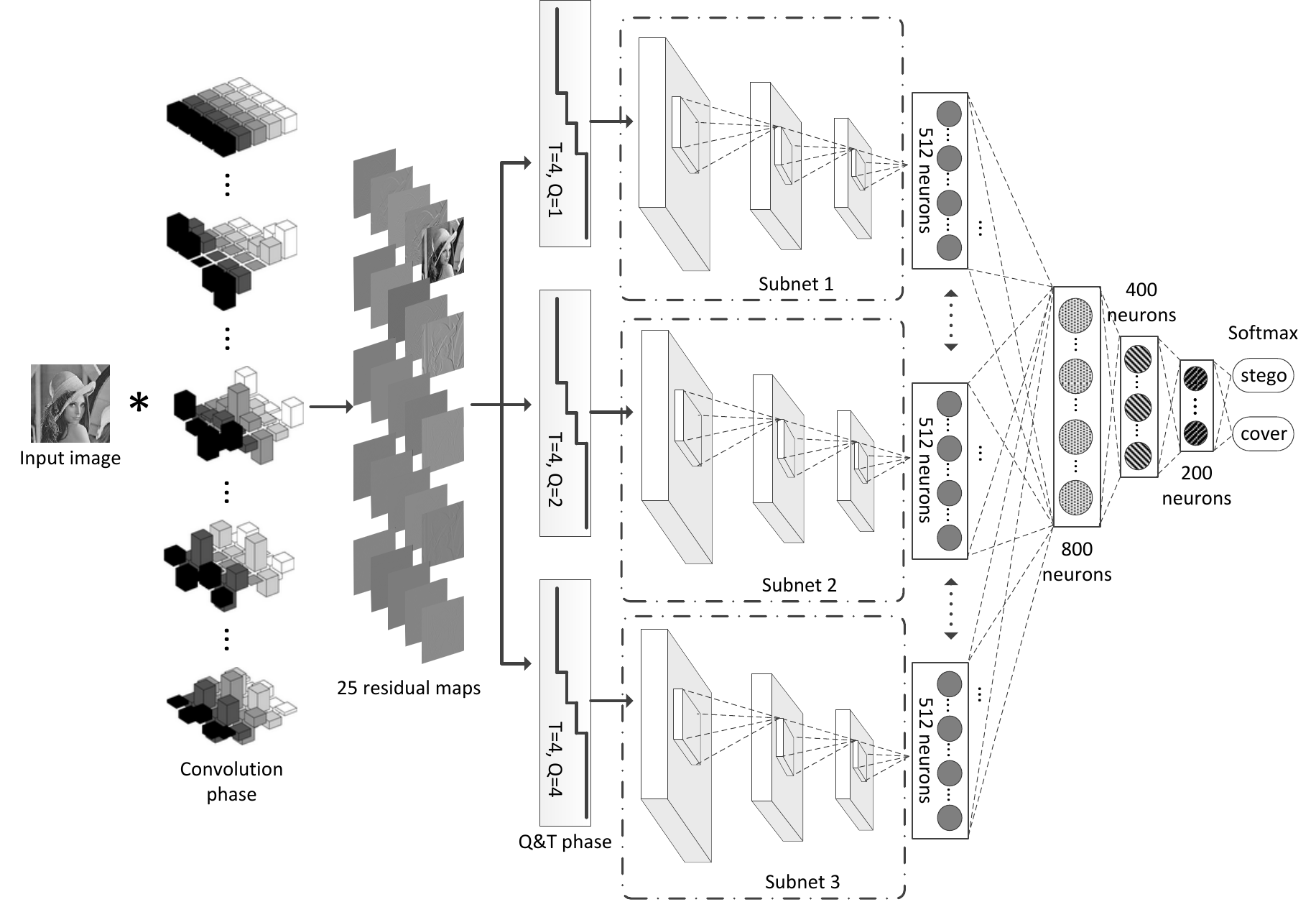}     
  \caption[]{Conceptual architecture of one implementation of our
    proposed hybrid deep-learning framework with twenty-five
    $5 \times 5$ DCT basis patterns and three Q\&T combinations. }
  \label{fig:1}
\end{figure*}

State-of-the-art rich models for JPEG
steganalysis~\cite{holub_tifs_2014,holub_spie_2015,song_ihmmsec_2015,denemark_tifs_2016}
take decompressed~(non-rounded and non-truncated) JPEG images as
input. The feature extraction procedure of JPEG rich models can be
divided into three phases:
\begin{itemize}
\item \textit{Convolution}: The target image is convolved with a set
  of kernels to generate diverse noise residuals. The purpose of this
  phase is to suppress the image contents as well as boost
  SNR~(Signal-to-Noise Ratio).
\item \textit{Quantization and truncation}~(Q\&T): Different quantized
  and truncated versions of each residual are calculated to further
  improve diversity of resulting features, as well as reduce the
  computational complexity.
\item \textit{Aggregation}: The values in noise residuals are
  aggregated to further reduce feature dimensionality.
\end{itemize}

Take DCTR~\cite{holub_tifs_2014} for example. Given a $M \times N$
JPEG image, it is firstly decompressed to the corresponding
spatial-domain version
$\textrm{\textbf{X}} \in \mathbb{R}^{M \times N}$. Sixty-four
$8 \times 8$ DCT basis patterns are defined as
$\textrm{\textbf{B}}^{(k,l)}=(B_{mn}^{(k,l)}), 0 \le k,l \le 7, 0 \le
m,n \le 7$:
\begin{equation}
  \label{eq:2.1-1}
  B_{mn}^{(k,l)}=\frac{w_kw_l}{4}cos\frac{\pi k(2m+1)}{16}cos\frac{\pi l(2n+1)}{16},
\end{equation}
where $w_0=\frac{1}{\sqrt{2}}$, $w_k=1$ for $k > 0$.
$\textrm{\textbf{X}}$ is convolved with $\textrm{\textbf{B}}^{(k,l)}$
to generate 64 noise residuals
$\textrm{\textbf{U}}^{(k,l)}, 0 \le k,l \le 7$:
\begin{equation}
  \label{eq:2.1-2}
  \textrm{\textbf{U}}^{(k,l)}=\textrm{\textbf{X}} \ast \textrm{\textbf{B}}^{(k,l)},
\end{equation}
Then the elements in each $\textrm{\textbf{U}}^{(k,l)}$ are quantized
with quantization step $q$ and truncated to a threshold $T$. The DCTR
features are constructed based on certain aggregation operation that
collect specific first-order statistics of the absolute values of the
quantized and truncated elements in each
$\textrm{\textbf{U}}^{(k,l)}$.

In \cite{tan_apsipa_2014}, we pointed out that in general the above
structure of rich models resembles CNN. Quantization and truncation
has become an indispensable part of rich steganalytic
models~\cite{fridrich_tifs_2012, denemark_wifs_2014,tang_tifs_2016,
  holub_tifs_2014,holub_spie_2015,
  song_ihmmsec_2015,denemark_tifs_2016}. However, as far as we know,
there still has been no published works regarding to the integration
of quantization and truncation into deep-learning steganalyzers.

In this paper, we would like to utilize the domain knowledge behind
rich models, especially the specific kernel matrices in the
convolutional phase and the Q\&T phase. But, The introduction of
quantization and truncation, namely the Q\&T phase on top of the
bottom convolution phase, is a double-edged sword. It cannot be put in the pipeline of
gradient-descent-based learning. The Q\&T phase takes noise residuals
generated by convolution phase as input, and can be modeled as:
\begin{equation}
  \label{eq:2.1-qt}
  a_{j}^{(2)}=f(z_j^{(2)})= \left\{
    \begin{array}{ll}
      \min([z_{j}^{(2)}/q],T) & \textrm{if}\ z_{j}^{(2)}>=0 \\
      \max([z_{j}^{(2)}/q],-T) & \textrm{if}\ z_{j}^{(2)}<0 \\
    \end{array}
  \right.
\end{equation}
where $z_{j}^{(2)}$ is an element of a given noise residual generated
by the bottom convolution phase, $a_{j}^{(2)}$ is the corresponding
activation output, $q$ is the quantization step, $[\cdot]$ denotes the
rounding operation, and $T$ is a predefined threshold. It is obvious
that $f'(z_i^{(2)})$ is zero along the entire domain of $z_{j}^{(2)}$
except the set of points
$\{(-T+0.5)q,(-T+1.5)q,\cdots,(T-1.5)q,(T-0.5)q\}$ where it is
infinite. Therefore \eqref{eq:2.1-qt} cannot be put in the pipeline of
gradient descent, since the derivative it passes on in backpropagation
will vanish. More specifically, the derivative does not exist if
$z_{j}^{(2)}$ is located at one of the points in the set
$\{(-T+0.5)q,(-T+1.5)q,\cdots,(T-1.5)q,(T-0.5)q\}$, otherwise the
derivative is equal to zero. The corresponding gradient saturates if
the partial derivative it passes on approaches to zero, and is
nullified if there is no derivative.

Incompatibility between Q\&T phase and gradient-descent-based learning
presents a dilemma in the design of deep-learning steganalytic
framework. The introduction of Q\&T phase implies that gradient
descent cannot be back propagated to the bottom convolution phase
without the usage of some unconventional bypass trick. The generic
hybrid deep-learning framework for JPEG steganalysis proposed in this
paper is intended to provide a solution to this dilemma.

\subsection{Our proposed hybrid deep-learning framework}
\label{sec:proposed_framework}

\begin{figure*}[!t]
  \centering
  \subfloat[]{
    \label{fig:2a}
    \includegraphics[width=0.25\linewidth,keepaspectratio]{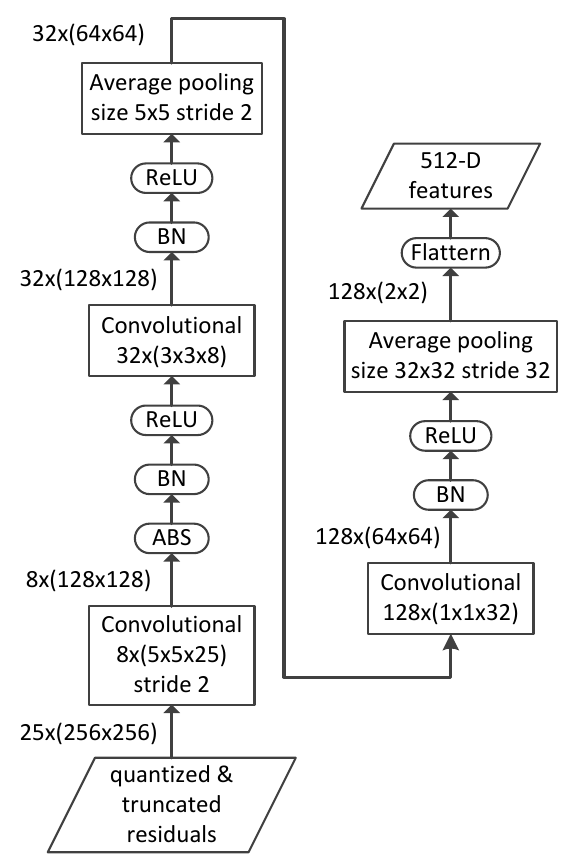}
  }\hspace{.2\textwidth}
  \subfloat[]{
    \label{fig:2b}
    \includegraphics[width=0.25\linewidth,keepaspectratio]{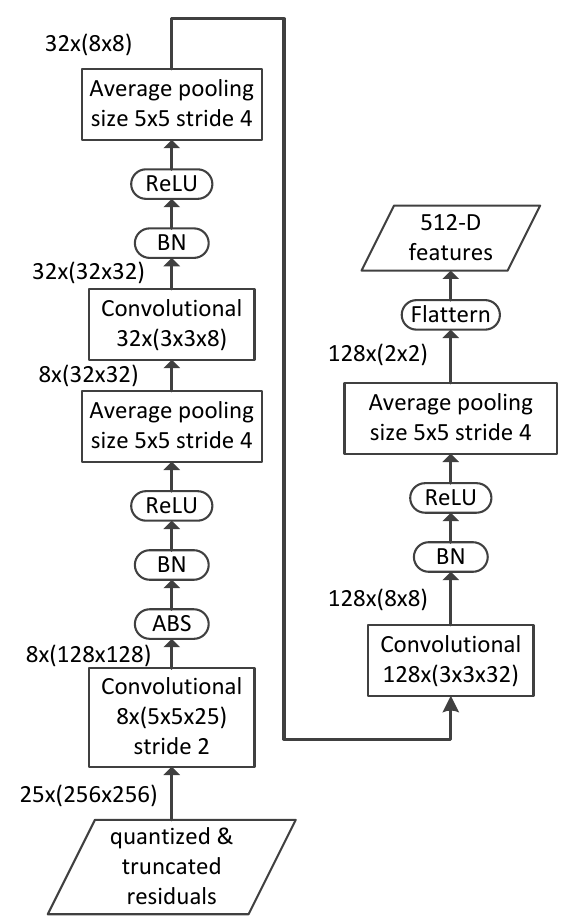} 
  }
  \caption[]{Two types of subnet
    configurations. \subref{fig:2a}~\textit{Type1}
    subnet. \subref{fig:2b}~\textit{Type2} subnet. In the two figures
    ``ABS'' denotes the activation layer which outputs absolute values
    of the corresponding inputs, ``BN'' denotes the batch
    normalization layer, and ``ReLU'' denotes the layer with
    rectified-linear-unit activation functions.}
  \label{fig:2}
\end{figure*}

Our proposed generic framework is composed of two stages. The first
stage takes decompressed~(non-rounded and non-truncated) JPEG images
as input, and corresponds to the convolution phase and the Q\&T phase
of rich models. The proposed generic framework can be implemented in
different way. The conceptual architecture of one implementation with
twenty-five $5 \times 5$ DCT basis patterns and three Q\&T
combinations is illustrated in Fig.~\ref{fig:1}.  In this
implementation, the first stage incorporated the first two phases of
DCTR~\cite{holub_tifs_2014}. All model parameters in this stage are
hand-crafted and gradient-descent-based learning is disabled. What
makes this stage different from DCTR is that DCTR uses sixty-four
$8 \times 8$ DCT basis patterns and only one Q\&T combination, while
our proposed approach contains twenty-five $5 \times 5$ DCT basis
patterns which are defined as
$\textrm{\textbf{B}}^{(k,l)}=(B_{mn}^{(k,l)}), 0 \le k,l \le 5, 0 \le
m,n \le 5$:
\begin{multline}
  \label{eq:2.2-1}
  B_{mn}^{(k,l)}=\frac{w_kw_l}{5}cos\frac{\pi k(2m+1)}{10}cos\frac{\pi
    l(2n+1)}{10},\\
  w_0=1,\ w_k=\sqrt{2}\ \textrm{for}\ k > 0.
\end{multline}
and three Q\&T combinations, namely $(T=4, Q=1)$, $(T=4, Q=2)$ and
$(T=4, Q=4)$. Given an input image, the convolution phase outputs
twenty-five residual maps. The residual maps pass through the Q\&T
phase. Three different groups of quantized and truncated residual maps
are generated. They constitute the input of the second stage. The
intention behind the design of the first stage of our proposed
framework is that we would like to utilize the domain knowledge behind
rich models, especially the specific kernel matrices in the
convolutional phase and the Q\&T phase. We agree with the concepts in
rich models~\cite{fridrich_tifs_2012}: model diversity is crucial to
the performance of steganalytic detectors. The model diversity of our
proposed framework is represented in twenty-five DCT basis patterns in
the hand-crafted convolutional layer and the three Q\&T combinations
that followed. There are total $25 \times 3=75$ sub-models in our
proposed framework.

The second stage is a compound deep CNN network in which the model
parameters are learned in the training procedure. The bottom of the
second stage is composed of three independent subnets with identical
structure. Each subnet corresponds to one group of quantized and
truncated residual maps. They take the residual maps as input and
generate three feature vectors. As shown in Fig.~\ref{fig:2}, within
this implementation, two types of subnet configurations are adopted.
Both of them contain three convolutional layers and output a
512-D~(512 dimensional) feature vector. \textit{Type1}
subnet~(Fig.~\ref{fig:2}\subref{fig:2a}) adopts $1 \times 1$
convolutional kernels in the top-most convolutional layer and uses a
single average pooling layer with large $32 \times 32$ pooling windows
at the end, as suggested in Xu's model~\cite{xu_spl_2016}. However,
deviated from the recipe suggested in Xu's model~\cite{xu_spl_2016}
that using TanH~(Hyperbolic Tangent) activation function in the lower
part, we always use ReLU~(Rectified Linear Unit) activation function
in \textit{Type1} subnet.  \textit{Type2}
subnet~(Fig.~\ref{fig:2}\subref{fig:2b}) is a traditional CNN
configuration. Compared with \textit{Type1} subnet, it adopts
progressive pooling layers and uses $3 \times 3$ convolutional kernels
in the top-most convolutional layer. Due to the progressing pooling
layers, \textit{Type2} subnet is a relative GPU memory-efficient
model. The GPU memory requirement of \textit{Type2} subnet is only
one-seventh of that of \textit{Type1} subnet. Both configurations have
in common are the BN layers which follow every convolutional layer.

In this implementation, three 512-D feature vectors output by the
bottom subnets are concatenated together to generate a single 1536-D
feature vector. The feature vector is subsequently fed into a
four-layer fully-connected neural network which makes the final
prediction. The successive layers of the fully-connected network
contain 800, 400, 200, and 2 neurons, respectively. ReLU activation
functions are used in all three hidden layers. The final layer
contains two neurons which denote ``stego'' prediction and ``cover''
prediction, respectively. Softmax function is used to output predicted
probabilities.

Recent researches on deep-learning revealed that ensemble prediction
with independently trained deep-learning models can improve the
performance~\cite{szegedy_cvpr_2015}. In \cite{xu_ihmmsec_2016}, Xu et
al. also demonstrated the potential of ensemble prediction in
deep-learning based steganalysis. Therefore, when compared to state of
the art in Sect.~\ref{sec:exp_compare}, we also introduce model ensemble in
the final prediction in order to further promote the detection
performance. Different from the approaches in \cite{xu_ihmmsec_2016},
we adopt a simple ensemble strategy, like the one used in
\cite{szegedy_cvpr_2015}. Five versions of our proposed deep-learning
models are independently trained with the same learning setting and
training dataset. They differ only in initial weights of the learnable
stage.  When testing, the decision of the five models are combined
with majority voting.

There is significant difference between our proposed framework and
other existing deep-learning steganalyzers~\cite{tan_apsipa_2014,
  qian_spie_2015,qian_icip_2016,pibre_ei_2016_ver2,xu_spl_2016,
  xu_ihmmsec_2016,sedighi_ei_2017}. Firstly, we explicitly introduce
the Q\&T phase used in rich models into our proposed deep-learning
steganalytic framework, which have never been seen in previous works.
Secondly, we adopt an array of dozens of hand-crafted convolutional
kernels in the bottom layer of our proposed framework, instead of an
image pre-processing layer with a single high-pass filter used in
previous works. And finally, there are three parallel CNN subnets with
identical structure in the central portion of our proposed framework,
which also have never been seen in previous works.

Our large-scale experiments reported in the following
Sect.~\ref{sec:exp} demonstrated that the introduction of Q\&T phase do
bring substantial detection performance improvement. The performance
improvement is not only due to the model diversity brought by
different Q\&T combinations~(as shown in
Sect.~\ref{sec:exp_impact_of_archit}). The discretization brought by
quantization and truncation itself also has an obvious impact on the
detection performance. We report the following experimental evidences
to support our argument. The experiments were conducted on basic500K
with setups shown in Sect.~\ref{sec:exp_setup}. J-UNIWARD stego images with
0.4bpnzAC~(bits per non-zero cover AC DCT coefficient) were included
in the experiments. In the experiments our proposed framework was
equipped with \textit{Type1} subnet. A corresponding model was trained
and tested independently for each configuration combination. We tested
the trained model every $10,000$ iterations, and reported the best
testing accuracy in $20 \times 10^4$ iterations. No ensemble
prediction was involved in this experiment, as in
Sect.~\ref{sec:exp_impact_of_archit}. The basic evidences are listed
as follows:
\begin{itemize}
\item The detection accuracy of Xu's model~\cite{xu_spl_2016}, which
  is without Q\&T phase, is merely \underline{54.7\%}.
\item The detection accuracy of our proposed framework as illustrated
  in Fig.~\ref{fig:1} is \underline{74.5\%}.
\item The detection accuracy of our proposed framework without Q\&T
  phase is \underline{61.5\%}.
\item The detection accuracy of our proposed framework without
  quantization step in the Q\&T phase, is \underline{57.6\%}, even
  worse than the above one without the entire Q\&T phase.
\item The detection accuracy of our proposed framework without
  truncation step in the Q\&T phase, is \underline{65.4\%}.
\end{itemize}
From the above experimental evidences we can clearly see that both
quantization and truncation effectively improve the detection
performance. 

As mentioned in the last section, the introduction of Q\&T phase
implies that gradient descent cannot be back propagated to the bottom
convolution phase. However, we still can back-propagate a fixed fake
tiny derivative $d$ to the bottom convolution phase.~\footnote{In
  practice, fake partial derivative can be back propagated to bottom
  layers when the actual partial derivative vanishes. For example,
  this trick is used in the Caffe implementation of ReLU
  layer~(\url{https://github.com/BVLC/caffe/blob/master/src/caffe/layers/relu_layer.cpp}).}
However, our extensive experiments show that such a fake derivative
just leads to serious performance degradation. For example, using a
Q\&T phase with a fixed fake derivative $d$, the detection accuracy of
our proposed framework as illustrated in Fig.~\ref{fig:1} is merely
\underline{60.5\%} when $d=0.01$, and \underline{56.8\%} when
$d=0.001$. Therefore, at present no compromise solution to the
incompatibility can be found.

But, does gradient-descent optimization of the bottom convolution
phase really matter? The following two experimental evidences reveal
that gradient-descent optimization of the bottom convolution phase
cannot improve the detection performance:
\begin{itemize}
\item The detection accuracy of Xu's model with a learnable bottom
  convolutional kernel, which is initialized as the high-pass filter
  used in \cite{xu_spl_2016}, is \underline{54.6\%}. Its performance
  is slightly worse than the one with fixed high-pass filter.
\item The detection accuracy of our proposed framework without Q\&T
  phase is \underline{61.3\%}, under the condition that
  gradient-descent- based learning is enabled for the bottom
  convolution phase. Its performance is also slightly worse than the
  one with fixed DCT basis patterns.
\end{itemize}

Recently in a similar field, image forensics, Bayar et al. proposed a
convolutional-layer regularizer which was claimed can be used to
suppress the content of an image~\cite{bayar_ihmmsec_2016}. However,
we observed that regularizing the bottom convolutional kernels using
the approach in \cite{bayar_ihmmsec_2016} did not lead to positive
changes in the above two experimental evidences:
\begin{itemize}
\item The detection accuracy of Xu's model is still
  \underline{54.6\%}.
\item The detection accuracy of our proposed framework without Q\&T
  phase is \underline{61.2\%}, slightly worse than the prior one.
\end{itemize}

All of the above experimental evidences reveal that at least in the
field of JPEG steganalysis, it is extraordinary difficult for an
existing deep-learning steganalytic framework to benefit from
gradient-descent optimization of the bottom convolution phase, under
the premise that the kernels in the bottom convolution phase have
already possessed the same parameters as those used in rich models. We
attribute this difficulty to the contradiction between the design
philosophy~(or domain knowledge) of the kernels in rich models and the
gradient descent algorithm used in deep-learning frameworks. The long
and widely accepted philosophy behind rich steganalytic models is that
high-pass kernels should be designed to extract the noise component
(noise residual) of images rather than their
content~\cite{fridrich_tifs_2012}. However, as shown in the
theoretical reflection in Appendix~\ref{app:reflection}, for a
deep-learning framework, we argue that the optimization of the bottom
convolutional kernels in favor of the extraction of stego noises is
hard to achieve with gradient descent.

The experimental demonstration in this section indicates that the
introduction of Q\&T phase do bring substantial detection performance
improvement. Certainly, the introduction of Q\&T phase is with
negative side effect: it blocks the back-propagated gradients. But the
theoretical reflection in Appendix~\ref{app:reflection} shows that
such negative side effect can be ignored, since even not cut off by
Q\&T phase, the back-propagated gradients is still hard to properly
guide the optimization of the bottom convolutional layer, as long as
its optimization goal is to benefit the extraction of stego noises. In
fact, the authors believe that we cannot directly draw on the design
philosophy of rich models to understand the underlying mechanism of
deep-learning steganalytic framework. Deep-learning frameworks are
trained and optimized as a whole. It may be not suitable to isolate
one part of a given deep-learning framework~(e.g. the learnable bottom
convolutional layer) and force it to comply with existing design
philosophy. Therefore our proposed hybrid deep-learning framework for
JPEG steganalysis is designed to be composed of two stages. The bottom
hand-crafted stage, which contains the convolution phase and the Q\&T
phase incorporated from rich models and complied with its design
philosophy, is not involved in gradient-descent-based optimization.
The second stage is a compound deep CNN network which does not need to
comply with the design philosophy of rich models, and is free to be
optimized using backpropagation as a whole.

\section{Experimental results}
\label{sec:exp}

\subsection{Experiment setups}
\label{sec:exp_setup}

We adopted ImageNet~\cite{imagenet_website}, a large-scale image
dataset containing more than fourteen million JPEG images, to evaluate
the steganalytic performance of our proposed hybrid deep-learning
framework. All of the experiments were conducted on a GPU cluster with
eight NVIDIA$^{\textrm{\textregistered}}$
Tesla$^{\textrm{\textregistered}}$ K80 dual-GPU cards. Independent
models are trained and tested in parallel, each of which is assigned
one GPU. By considering the computation capacity, we restricted the
size of the target images to $256 \times 256$. We randomly selected 50
thousand, 500 thousand and 5,000 thousand~(namely 5 million) JPEG
images with size larger than $256 \times 256$ from ImageNet. Their
left-top $256 \times 256$ regions were cropped, converted to grayscale
and then re-compressed as JPEG with quality factor 75.\footnote{The
  original quality factors of ImageNet images are diverse. Out of 10
  million ImageNet images with size larger than $256 \times 256$,
  there are more than 1.5 million images whose quality factors cannot
  be detected by ImageMagick utility ``identify'', and roughly 8.3
  million images with diverse quality factors which are larger than
  75. We uniformly converted the quality factors of the selected
  images to 75 due to the following two reasons: Firstly, all the
  reported experiments of previous works, including DCTR, PHARM, GFR,
  and SCA, are conducted on images with quality factor 75 and 95. And
  secondly, if the target quality factor is set to 95, then for a
  majority of the selected images, we need to elevate their quality
  factors which may introduce exploitable artifacts.} The resulting
images constituted the following three basic cover image datasets:
\begin{itemize}
\item basic50K: The small-scale dataset used in our experiments. By
  comparing the detection performance of our proposed framework on
  basic50K and basic500K~(see below), we can highlight the superiority
  of our proposed framework in large-scale dataset.
\item basic500K: The major dataset for most all of our experiments,
  including the verification experiments to determine hyper-parameters
  of our proposed framework.
\item basic5000K: The largest-scale dataset used in our
  experiments. Due to the limitation of computation capacity, we only
  conducted the experiments on stego images with 0.4 bpnzAC.
\end{itemize} 

\begin{figure*}[!t]
  \centering
  \subfloat[]{
    \label{fig:4a}
    \includegraphics[width=0.4\linewidth,keepaspectratio]{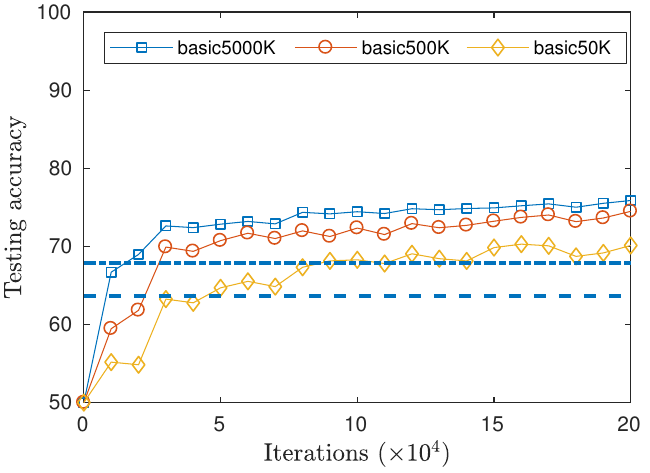} 
  }
  \subfloat[]{
    \label{fig:4b}
    \includegraphics[width=0.4\linewidth,keepaspectratio]{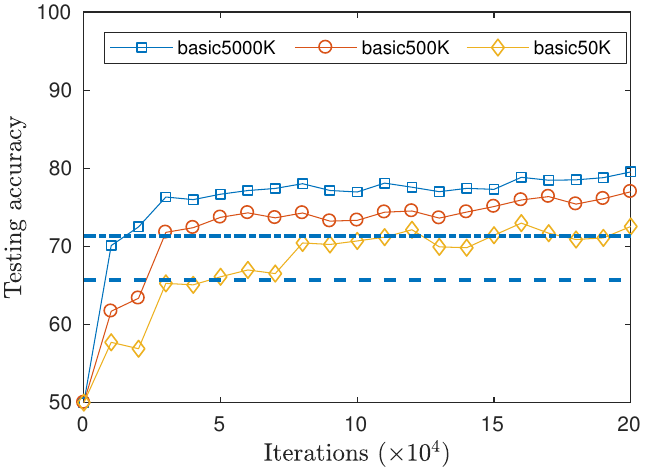}
  }
  \caption[]{Testing accuracies versus training iterations for our
    proposed framework. The experimental results on basic50K,
    basic500K and basic5000K are reported. For brevity, only stego
    images with 0.4bpnzAC were included in the
    experiments. \subref{fig:4a} is for J-UNIWARD steganography while
    \subref{fig:4b} is for UERD steganography. In \subref{fig:4a} and
    \subref{fig:4b}, The dash-dotted and the dashed reference lines
    denote the best testing accuracy of GFR and DCTR in basic500K,
    respectively.}
  \label{fig:4}
\end{figure*}

Our implementation was based on the publicly available Caffe
toolbox~\cite{jia_arxiv_2014} with our implemented hand-crafted
convolutional layer~(with $5 \times 5$ DCT basis patterns) and Q\&T
layer according to \eqref{eq:2.1-qt}. Our proposed models were trained
using mini-batch stochastic gradient descent with ``step'' learning
rate starting from 0.001~(stepsize: 5000; weight\_decay: 0.0005;
gamma: 0.9) and a momentum fixed to 0.9. The batch size in the
training procedure was 64 and the maximum number of iterations was set
to $20 \times 10^4$. In each experiment, we tested the trained model
in the corresponding standalone testing set every $10,000$ iterations,
and reported the best testing accuracy in $20 \times 10^4$ iterations.
Please note that as later shown in Fig.~\ref{fig:4}, when trained on a
large-scale dataset such as basic500K, our proposed framework
exhibited good convergence and stability after less than $5 \times
10^4$ iterations. Therefore validation set was omitted for the sake of
resources saving. The source code and auxiliary materials are
available for download from
GitHub~\footnote{\url{https://github.com/tansq/hybrid_deep_learning_framework_for_jpeg_steganalysis}}.

J-UNIWARD~\cite{holub_eurasip_2014}, UERD~\cite{guo_tifs_2015} and
UED~\cite{guo_tifs_2014}, the three state-of-the-art JPEG
steganographic schemes, were our attacking targets in the experiments.
The default parameters of the three steganographic schemes were
adopted in our experiments. 50\% cover images were randomly selected
from basic50K, basic500K, and basic5000K, respectively. They
constituted the training set along with their corresponding stego
images. The rest 50\% cover-stego pairs in the dataset were for
testing. We further guaranteed that the cover images included in an
arbitrary training set of the three datasets would not appear in any
of the three testing sets.
\subsection{Impact of the framework architecture on the performance}
\label{sec:exp_impact_of_archit}

In Tab.~\ref{tab:1}, we compare the effect of different Q\&T
combinations, different hand-crafted convolutional kernels, and the
presence of BN layers. The experiment was conducted on basic500K. A
corresponding model is trained and tested independently for each
configuration combination. No ensemble prediction is involved in this
experiment. We can see that under the same conditions, DCT basis
patterns~(including the $8 \times 8$ DCTR
kernels~\cite{holub_tifs_2014}) always perform better than PHARM
kernels~\cite{holub_spie_2015}. The experimental results support our
choice of DCT basis patterns. $5 \times 5$ DCT basis patterns can
achieve significant performance improvement compared to $3 \times 3$
DCT basis patterns. However, the performance of the more complex
$8 \times 8$ DCTR kernels is not even as good as the $3 \times 3$ DCT
basis patterns, which indicates that increasing the size of the
convolutional kernels is not always beneficial at the cost of
increasing model complexity. The performance of GFR
kernels~\cite{song_ihmmsec_2015} is slightly better than $5 \times 5$
DCT basis patterns. However, with as many as two hundred and fifty-six
output residual maps, GFR kernels are too resource consuming to be
included in our proposed framework. Different Q\&T combinations also
affect the performance of our proposed framework. Combinations with
three different quantization steps and the same threshold are of
relatively cost-effective. BN layers in the subnets are crucial,
especially the first one at the bottom of the subnets. Therefore,
based on the described results, we adopt twenty-five $5 \times 5$ DCT
basis patterns, $T=4, Q=[1,2,4]$ and subnet configurations with a BN
layer following every convolutional layer in our final proposed
framework.

\begin{figure*}[!t]
  \centering
  \subfloat[]{
    \label{fig:3a}
    \includegraphics[width=0.4\textwidth,keepaspectratio]{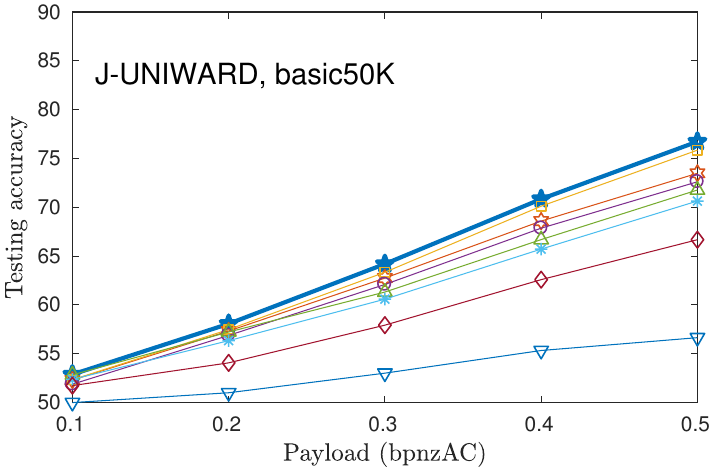} 
  }
  \subfloat[]{
    \label{fig:3b}
    \includegraphics[width=0.4\textwidth,keepaspectratio]{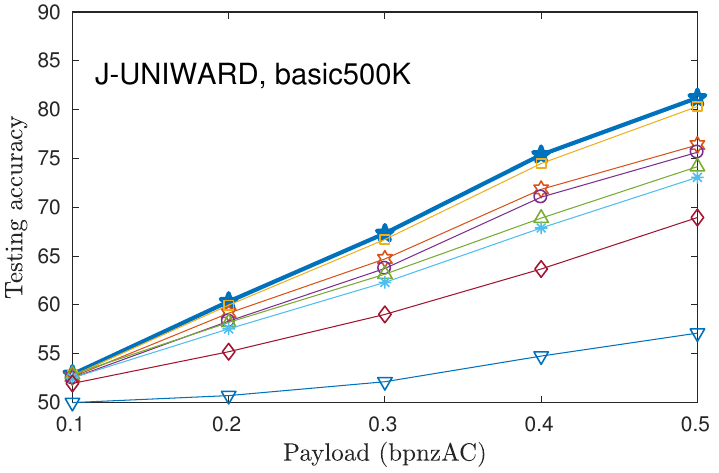}
  }\\
  \subfloat[]{
    \label{fig:3c}
    \includegraphics[width=0.4\textwidth,keepaspectratio]{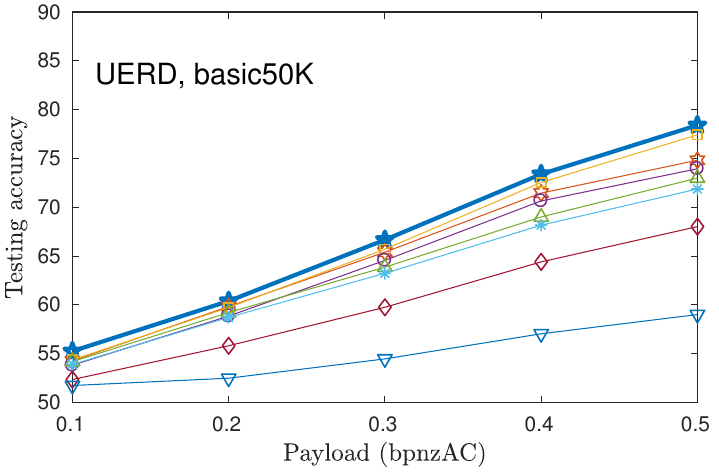}
  }
  \subfloat[]{
    \label{fig:3d}
    \includegraphics[width=0.4\textwidth,keepaspectratio]{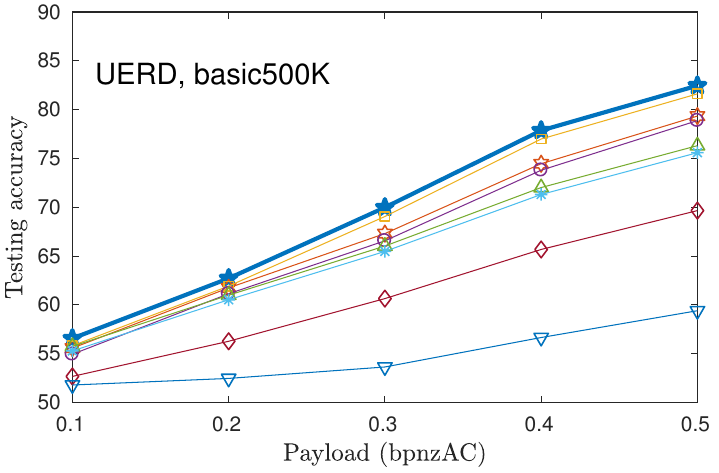}
  }\\
  \subfloat[]{
    \label{fig:3e}
    \includegraphics[width=0.4\textwidth,keepaspectratio]{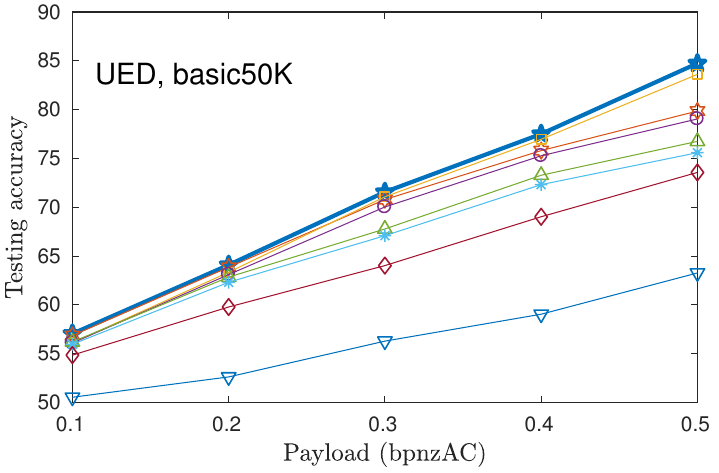}
  }
  \subfloat[]{
    \label{fig:3f}
    \includegraphics[width=0.4\textwidth,keepaspectratio]{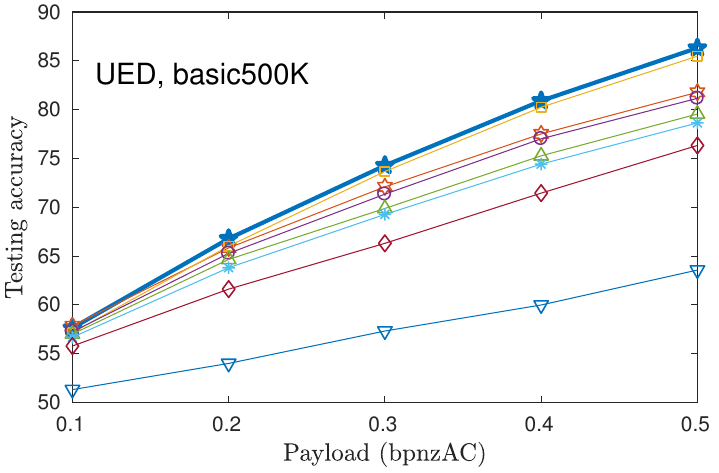}  
  }\\
  \includegraphics[scale=0.7]{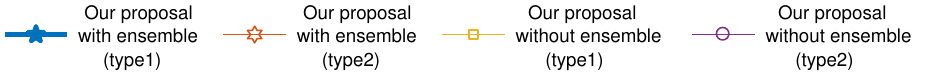}\\
  \includegraphics[scale=0.7]{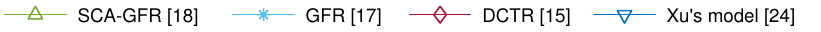} 
  \caption[]{Comparison of testing accuracy of our proposed frameworks
    with four steganalytic models described in the literature, two
    hand-crafted JPEG domain rich models~(DCTR and GFR), a
    selection-channel-aware variant of GFR~(SCA-GFR) and a
    deep-learning steganalytic model proposed by Xu et
    al.~\cite{xu_spl_2016}. \subref{fig:3a} and \subref{fig:3b} are
    the results for J-UNIWARD; \subref{fig:3c} and \subref{fig:3d} are
    for UERD; \subref{fig:3e} and \subref{fig:3f} are for UED. The
    experiments for \subref{fig:3a}, \subref{fig:3c} and
    \subref{fig:3e} were conducted on basic50K, while those for
    \subref{fig:3b}, \subref{fig:3d} and \subref{fig:3f} were
    conducted on basic500K.}
  \label{fig:3}
\end{figure*}

\subsection{Comparison to state of the art}  
\label{sec:exp_compare}

In Fig.~\ref{fig:3}, we compare the performance of our proposed
framework and other steganalytic models in the literature.  Please
note that for a fair comparison, Xu's model~\cite{xu_spl_2016} is also
fed with decompressed~(non-rounded and non-truncated) images, and is
trained with the same learning protocol as that for our own
model~\footnote{The original Xu's model is fed with $512 \times 512$
  images. In order to make it adapt to $256 \times 256$ inputs used in
  our experiments, we explicitly set ``stride=2'' for its bottom
  convolutional layer which takes the residual map generated by the KV
  kernel as input. Please note that we also set ``stride=2'' in the
  bottom convolutional layer of \textit{Type1} and \textit{Type2}
  subnets of our proposed framework.}.  From Fig.~\ref{fig:3} we can
see that our proposed framework can obtain significant performance
improvement compared with DCTR~\cite{holub_tifs_2014},
GFR~\cite{holub_spie_2015}, and even recently proposed
selection-channel-aware JPEG rich model
SCA-GFR~\cite{denemark_tifs_2016}. For all of the three steganographic
algorithms, the performance of Xu's model was unsatisfactory. The
degraded performance of Xu's model is acceptable, since it is designed
for spatial-domain steganalysis.  The superiority of our proposed
framework is more obvious in basic500K.  This is due to the fact that
with more training samples raised by one magnitude, the large-scale
basic500K dataset with 500,000 training samples~(covers plus the
corresponding stegos) is more favor of deep-learning frameworks like
the one proposed by us. If only consider the performance of a single
model, our proposed framework with \textit{Type1} subnets behaved
better than its companion with \textit{Type2} subnets. Furthermore,
the final prediction conducted by the ensemble of five independently
trained models shows that model ensemble could improve the detection
accuracy by $1\%$ regardless of the type of the underlying subnet
configurations.\footnote{Please note that the ensemble approach of
  Xu's model~\cite{xu_spl_2016} can also probably obtain better
  results. The experimental results of ensemble prediction of Xu's
  model are omitted in Fig.~\ref{fig:3} for clarity.} Since the
performance of our proposed framework with \textit{Type1} subnets is
always better than that with \textit{Type2} subnets, we insisted on
using \textit{Type1} subnets in the following experiments. However,
please note that \textit{Type2} subnet can potentially be used in more
complex deep-learning steganalytic frameworks in the future since it
is a memory-efficient model.

\begin{table}[!t]
  \begin{minipage}{\columnwidth}
    \centering
    \caption[]{Effect of different Q\&T combinations, different
      hand-crafted convolutional kernels, and the presence of BN
      layers. Only J-UNIWARD stego images with 0.4bpnzAC were included
      in the experiments. The best results in every sub-table are
      underlined. Those hyper-parameters adopted in our proposed
      framework are marked in bold.\footnote{Logograms are used in
        expressing Q\&T combinations. For example, (4,1) denotes
        $(T=4, Q=1)$.} }
   \label{tab:1}
   \resizebox{\columnwidth}{!}{%
     {\renewcommand{\arraystretch}{1.2}
       \begin{tabular}{cccc}
         \Xhline{2\arrayrulewidth}
         \multirow{2}{9em}{Threshold \& Quantization Steps} &
         \multicolumn{3}{ c }{BN Layers} \\
         \cline{2-4}
         & \textbf{With BNs} & Without BN1 & Without BNs \\
         \hline
         \multicolumn{4}{ c }{Nine $3 \times 3$ DCT basis patterns} \\
         \hline
         (4,1), (4,1.5), (4,2) & 73.1\% & 70.6\% & 50.1\% \\
         (4,2), (4,2), (4,2) & 72.8\% & 70.1\% & 50.0\% \\
         \textbf{(4,1), (4,2), (4,4)} & \underline{73.2\%} & 71.0\% & 50.1\% \\
         (2,1), (4,2), (6,4) & 71.2\% & 68.5\% & 50.0\% \\
         (6,1), (4,2), (2,4) & 70.6\% & 67.8\% & 50.0\% \\
         \hline
         \multicolumn{4}{ c }{\textbf{Twenty-five $5
             \times 5$ DCT basis patterns}} \\
         \hline
         (4,1), (4,1.5), (4,2) & 74.3\% & 72.4\% & 50.1\% \\
         (4,2), (4,2), (4,2) & 74.1\% & 72.4\% & 50.1\% \\
         (4,1) & 70.8\% & 69.4\% & 50.1\% \\
         (4,1), (4,2) & 72.5\% & 70.2\% & 50.1\% \\  
         \textbf{(4,1), (4,2), (4,4)} & \underline{74.5\%} & 72.5\% &
         50.1\% \\
         (2,1), (4,2), (6,4) & 73.6\% & 72.0\% & 50.1\% \\
         (6,1), (4,2), (2,4) & 72.6\% & 71.7\% & 50.0\% \\
         \hline
         \multicolumn{4}{ c }{Sixty-four $8 \times 8$ DCTR
           kernels~\cite{holub_tifs_2014}} \\
         \hline
         (4,1), (4,1.5), (4,2) & 72.5\% & 71.4\% & 50.0\% \\
         (4,2), (4,2), (4,2) & 72.7\% & 71.2\% & 50.1\% \\
         \textbf{(4,1), (4,2), (4,4)} & \underline{72.9\%} & 71.2\% & 50.1\% \\
         (2,1), (4,2), (6,4) & 71.9\% & 70.2\% & 50.0\% \\
         (6,1), (4,2), (2,4) & 71.5\% & 70.1\% & 50.1\% \\
         \hline
         \multicolumn{4}{ c }{Thirty $5 \times 5$ PHARM kernels~\cite{holub_spie_2015}} \\
         \hline
         (4,1), (4,1.5), (4,2) & 72.0\% & 70.8\% & 50.1\% \\
         (4,2), (4,2), (4,2) & 70.6\% & 68.8\% & 50.0\% \\
         \textbf{(4,1), (4,2), (4,4)} & \underline{72.1\%} & 70.8\% & 50.1\% \\
         (2,1), (4,2), (6,4) & 70.3\% & 68.6\% & 50.0\% \\
         (6,1), (4,2), (2,4) & 70.2\% & 68.7\% & 50.0\% \\
         \hline
         \multicolumn{4}{ c }{Two hundred and fifty-six $8 \times 8$ GFR kernels~\cite{song_ihmmsec_2015}} \\
         \hline
         (4,1), (4,1.5), (4,2) & 74.1\% & 72.5\% & 50.1\% \\
         (4,2), (4,2), (4,2) & 74.0\% & 72.6\% & 50.1\% \\
         \textbf{(4,1), (4,2), (4,4)} & \underline{74.6\%} & 72.5\% &
         50.0\% \\
         (2,1), (4,2), (6,4) & 74.1\% & 72.4\% & 50.0\% \\
         (6,1), (4,2), (2,4) & 72.3\% & 71.5\% & 50.0\% \\
         \Xhline{2\arrayrulewidth}
       \end{tabular}
     }}
   \end{minipage}
\end{table}

In Fig.~\ref{fig:4} we show how the testing accuracy changes with
successive training iterations in the experiments which were conducted
on basic50K, basic500K and basic5000K, our largest-scale dataset,
respectively. The tests were performed on standalone testing dataset
every $10,000$ training iterations and the models were trained for
$20 \times 10^4$ iterations in total. Only stego images with 0.4bpnzAC
were included in the experiments due to the limited computational
capacity. Even so for basic5000K there were five million
images~(covers plus the corresponding stegos) involved in a training
epoch. Our proposed deep-learning framework showed strong learning
capacity that further improves along with the growth of training
samples. From Fig.~\ref{fig:4} we can also see that the curve of
testing accuracy for the framework trained on basic5000K not only is
of the best performance but also is of the best stability. Please note
that $20 \times 10^4$ iterations is roughly equivalent to 256 epochs
for basic50K, 25.6 epochs for bsic500K, and only 2.56 epochs for
basic5000K. Therefore the full potential of our proposed framework
with large-scale training datasets may not have been fully
exploited.\footnote{The implementation of ensemble
  classifier~\cite{kodovsky_tifs_2012} used by rich models cannot be
  scaled to large-scale datasets. Therefore we cannot provide the
  testing accuracy of DCTR and GFR in basic5000K for comparison in
  Fig.~\ref{fig:4}.}

Throughout the experiments, our proposed framework ran steadily.
During the training procedure, it could accomplish 1,000 iterations
every 20 minutes. That is to say, $20 \times 10^4$ training iterations
could be finished in about 67 hours. With K80 GPU cards, We can expect
to finish one epoch of training in 0.26 hour, 2.6 hours, and 26 hours
for basic50K, basic500K, and basic5000K,
respectively. 

\begin{figure}[!t]
  \centering
  \includegraphics[width=0.8\columnwidth,keepaspectratio]{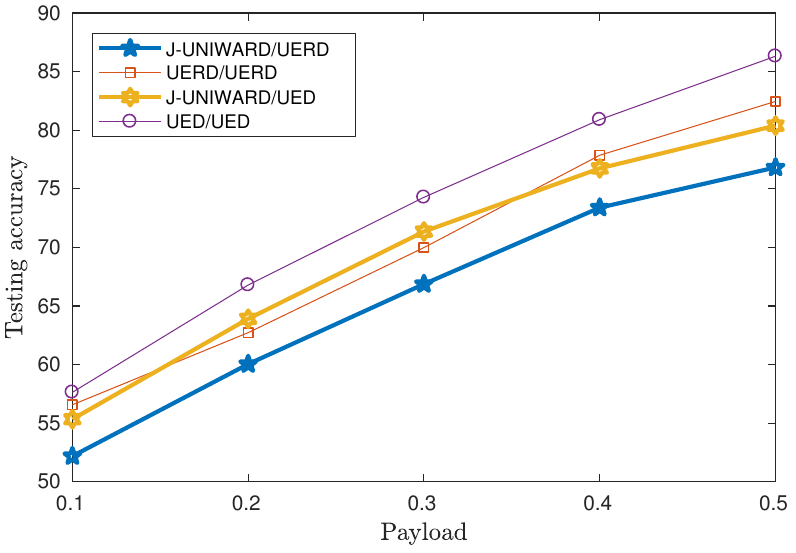}   
  \caption[]{Comparison of attacking-target transfer ability of our
    proposed framework. The experiments were conducted on basic500K
    dataset. Only stego images with 0.4bpnzAC were included in the
    experiments. The notations in the legend take the form of the
    target in training and the target in testing delimited by a
    slash~(/). For example, ``J-UNIWARD/UERD'' means that J-UNIWARD
    stego images were used in training while UERD stego images were
    used in testing.}
  \label{fig:attacking_target_transfer}
\end{figure} 

\begin{figure}[!t]
  \centering 
  \includegraphics[width=0.9\columnwidth,keepaspectratio]{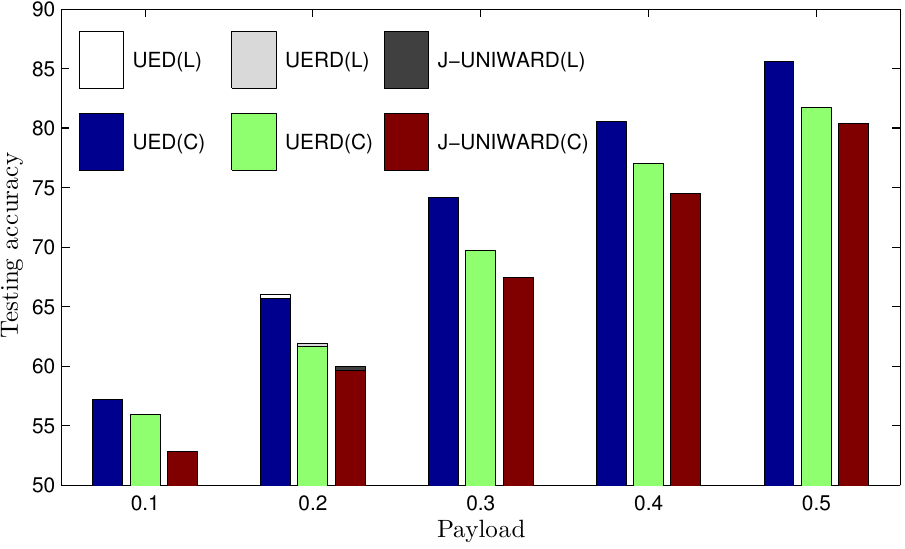}  
  \caption[]{The impact of altered blocking artifact on the
    performance of our proposed framework. Only stego images with
    0.4bpnzAC were included in the experiments. All of the models were
    trained on basic500K training set and then tested on the
    corresponding testing set with central-cropped images.  The legend
    ``C'' in parentheses denotes those tested on central-cropped
    images, while ``L'' in parentheses denotes those tested on the
    original basic500K testing set. For example, ``J-UNIWARD~(C)''
    means that the corresponding framework was trained and tested with
    J-UNIWARD stego images. It was trained on basic500K training set
    and then tested on the corresponding testing set with
    central-cropped images.}   
  \label{fig:altered_blocking_artifact}
\end{figure}

\begin{figure}[!t]
  \centering
  \includegraphics[width=0.8\columnwidth,keepaspectratio]{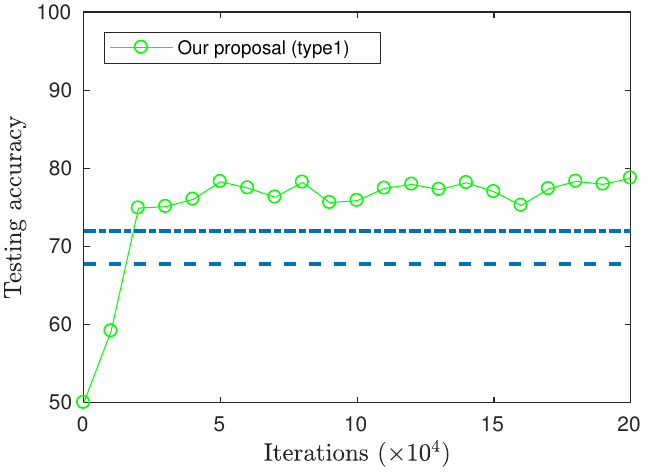}  
  \caption[]{Testing accuracies versus training iterations for our
    modified framework which takes $512 \times 512$ images as input.
    Only J-UNIWARD stego images with 0.4bpnzAC are included in the
    experiment.  As in Fig.~\ref{fig:4}, The dash-dotted reference line
    denotes the best testing accuracy of GFR, while the dashed
    reference line denotes the best testing accuracy of DCTR in the
    same testing dataset.}
  \label{fig:512x512}
\end{figure}

\begin{figure}[!t]
  \centering
  \includegraphics[width=0.8\columnwidth,keepaspectratio]{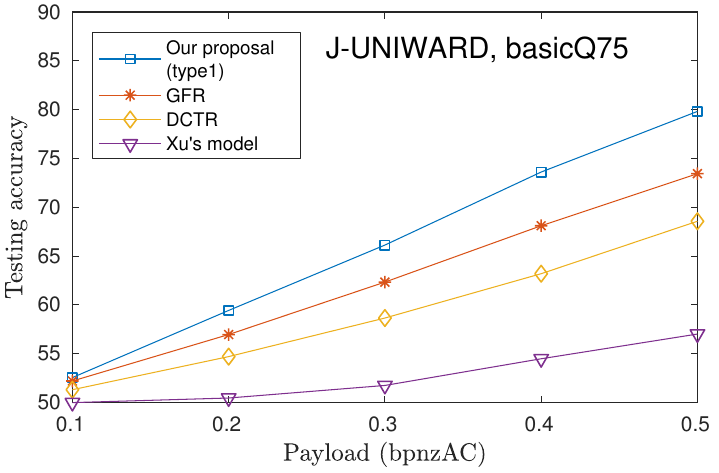}   
  \caption[]{Comparison of testing accuracy of our proposed framework
    with GFR, DCTR, and Xu's model for J-UNIWARD on basicQ75 dataset.}
  \label{fig:fig_results_basicq75}
\end{figure}  

\begin{figure}[!t]
  \centering
  \includegraphics[width=0.8\columnwidth,keepaspectratio]{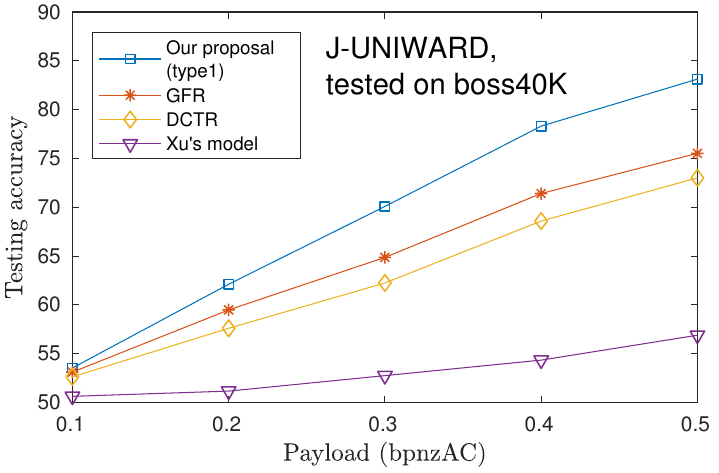}   
  \caption[]{Comparison of testing accuracy of our proposed framework
    with GFR, DCTR, and Xu's model for J-UNIWARD on boss40K dataset.}
  \label{fig:fig_results_boss40k}  
\end{figure}   

\begin{figure}[!t]
  \centering
  \includegraphics[width=0.8\columnwidth,keepaspectratio]{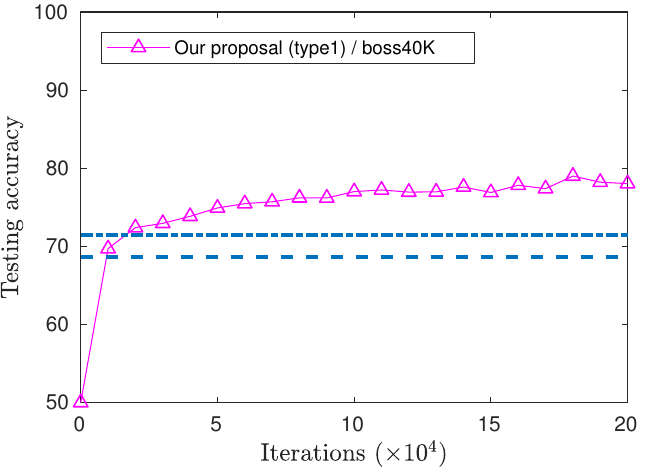} 
  \caption[]{Testing accuracies versus training iterations for our
    proposed framework. The models are trained on basic500K while
    tested on boss40K. Only J-UNIWARD stego images with 0.4bpnzAC are
    included in the experiment. The dash-dotted line and the dashed
    line denote the best testing accuracy of GFR and DCTR,
    respectively.}
  \label{fig:fig_acc_vs_iter_boss40k}
\end{figure} 

\begin{figure}[!t]
  \centering
  \includegraphics[width=0.8\columnwidth,keepaspectratio]{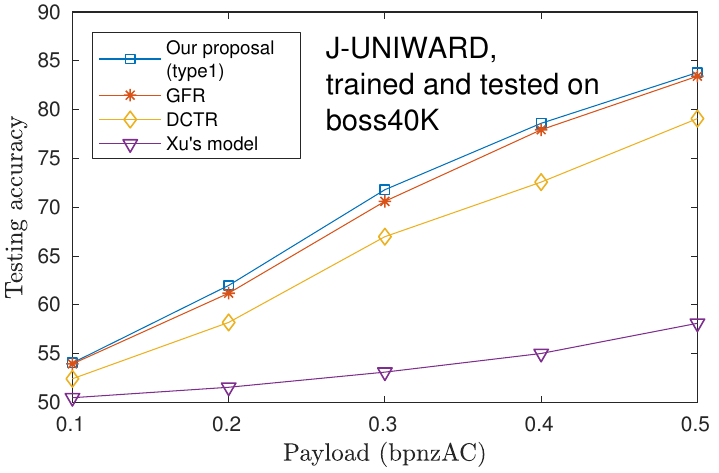}   
  \caption[]{Comparison of testing accuracy of our proposed framework
    with GFR, DCTR, and Xu's model for J-UNIWARD on boss40K dataset,
    when all of them were also trained on boss40K.}
  \label{fig:fig_results_train_test_all_in_boss40k_qf75} 
\end{figure}  

\subsection{Performance with mismatched targets, altered blocking
  artifact, doubled-sized inputs and single-compressed images}
\label{sec:exp_misc}

First of all, please note that in the following experiments, our
proposed framework is equipped with \textit{Type1} subnet. No ensemble
prediction is involved to reduce the time of experiments.

In Fig.~\ref{fig:attacking_target_transfer}, we observe the
attacking-target transfer ability of our proposed framework. The
framework was trained with J-UNIWARD cover/stego pairs and then tested
with UERD/UED cover/stego pairs. The detection accuracy is roughly
$3\%-4\%$ worse compared with that trained and tested with the same
type of stego images. However, the degradation of detection
performance is acceptable especially for the detection of UED stego
images given that UED works in a very different way compared with
J-UNIWARD.

$8 \times 8$ block processing during JPEG compression introduces
blocking artifacts, which can be used as intrinsic statistical
characteristic of JPEG cover images. Secret bits embedded in the DCT
domain tend to impair blocking artifacts, therefore leave traces which
can be utilized by steganalyzers. An interesting problem is to access
the performance of our proposed framework depending on the intrinsic
statistical characteristic of blocking artifacts. In
Fig.~\ref{fig:altered_blocking_artifact}, we observe the impact of
altered blocking artifacts on the performance of our proposed
framework. The default testing set in basic500K contains left-top
cropped images in which the original DCT grid alignment is preserved.
In this experiment for all the testing images in basic500K, we
re-compressed their corresponding original images in ImageNet with
quality factor 75 and then converted them to grayscale images
again. We cropped their central $256 \times 256$ regions to constitute
a new testing set. The motivation is that central cropping cannot
preserve the original DCT grid alignment in most cases, therefore the
blocking artifacts from two different sources coexist. As a result the
blocking artifacts in the images of the new testing set are different
from those in the training set. However,
Fig.~\ref{fig:altered_blocking_artifact} reveals that the impact of
altered blocking artifact on the performance of our proposed framework
is small. Our proposed framework has captured more complex intrinsic
statistical characteristic besides blocking artifact.

All of the above experiments used images of size $256 \times 256$
pixels. This limitation stems mainly from the following two factors:
Firstly, target images with larger size, e.g. $512 \times 512$ pixels
result in deep-learning models hard to train with K80 GPU cards we
have in hands. Secondly, large-sized ImageNet images are in the
minority. Out of fourteen million ImageNet images, only roughly 0.7
million of them are larger than $512 \times 512$ pixels. In the
following experiment, we tested our proposed framework with
double-sized inputs on this limited dataset. 500 thousand JPEG images
with size larger than $512 \times 512$ were randomly picked out from
ImageNet and were converted to $512 \times 512$ with the same
processing procedure as mentioned in Sect.~\ref{sec:exp_setup}. Due to GPU
memory constraints, we simplified the model by using doubled stride in
the convolutional layer of each subnet~(i.e. 4 instead of 2). All
other experiment setups were remained the same except that the batch
size in the training procedure is reduced to 32. Only J-UNIWARD stego
images with 0.4bpnzAC were included in the experiment.
Fig.~\ref{fig:512x512} shows the testing accuracy in successive
training iterations. The training procedure also converged quickly and
delivered better performance than the DCTR and GFR models. Due to the
limited computational capacity, subnets with wider and deeper
structures were not evaluated in this experiment. Its potential for
target images with larger size may have not been fully demonstrated.

Up to now we used double-compressed images in the experiments.  As
reported by Pibre et al.~\cite{pibre_ei_2016_ver2}, CNN based
steganalyzers can take advantage of seemingly irrelevant subtle
patterns to boost their performance. We must eliminate the possibility
that our proposed framework makes use of the double compression
artifacts to dispel the doubts of the colleagues. Hence, we conducted
two more experiments with single-compressed JPEG images.

Firstly, There are about $410,000$ ImageNet images can be confirmed as
being compressed with quality factor 75. They were all selected.
Their left-top $256 \times 256$ regions were cropped, converted to
grayscale without double compression to constitute a new dataset
``basicQ75''. $200,000$ cover images were randomly selected from them
for training while the rest were for testing. In
Fig.~\ref{fig:fig_results_basicq75}, we compare the performance of our
proposed framework with three other steganalyzers for J-UNIWARD on
bacicQ75 dataset. For the sake of brevity, only the results of half of
the steganalyzers listed in Fig.~\ref{fig:3} are listed in
Fig.~\ref{fig:fig_results_basicq75}.  However, by comparing
Fig.~\ref{fig:fig_results_basicq75} and
Fig.~\ref{fig:3}\subref{fig:3b}, we still can find that as with all
other three steganalyzers, our proposed framework only suffered slight
performance degradation, which may be attributed to the relative lack
of diversity in basicQ75 dataset.

Secondly, we divided every image in BOSSBase public
dataset~\cite{bas_ih_2011} into four equal parts and then JPEG
compressed them with quality factor 75.  Through this method, we
obtained $40,000$ single-compressed JPEG cover images. We denoted them
as ``boss40K'' dataset, and used all of the $40,000$ cover images and
the corresponding stego images to test the performance of our proposed
framework and other steganalyzers trained on basic500K. We prefer to
use all of the images in boss40K dataset in testing rather than in
training, which is based on the following two aspects: 1. Merely
$40,000$ images are not suitable for training a deep-learning
steganalyzer with hundreds of thousands of learnable parameters. 2. As
a dataset with totally different source, boss40K is more suitable for
checking transfer ability of steganalyzers trained with ImageNet
images.

In Fig.~\ref{fig:fig_results_boss40k}, we show the testing results of
our proposed framework~(with \textit{Type1} subnet, without ensemble)
and other steganalyzers on boss40K. Please note that all the
steganalyzers used in this experiment were trained on basic500K. By
comparing Fig.~\ref{fig:fig_results_boss40k} and
Fig.~\ref{fig:3}\subref{fig:3b}, we are delighted to find that our
proposed framework even achieved better detection performance, and its
superiority over all other three steganalyzers became more
obvious. Fig.~\ref{fig:fig_acc_vs_iter_boss40k} shows testing accuracy
of our proposed framework on boss40K in successive training
iterations. Please note that the model was also trained on
basic500K. From Fig.~\ref{fig:fig_acc_vs_iter_boss40k} we can see our
proposed framework trained on basic500K exhibited rapid convergence
even when evaluated on a dataset with totally different source, which
provides complementary evidence to support the removal of validation
set in our large-scale experiments.

For the sake of completeness, we also show the testing results of our
proposed framework~(with \textit{Type1} subnet, without ensemble) and
other steganalyzers on boss40K dataset, when all of them were also
trained on boss40K in
Fig.~\ref{fig:fig_results_train_test_all_in_boss40k_qf75}. Since
validation cannot be omitted for a small-scale dataset, boss40K was
split into 60/15/25 ratio, for training, validating, and testing,
respectively. We guaranteed that all the sub-images of a given
BOSSBase image could only be assigned to one sub-dataset. Please note
that our proposed framework aims at large-scale JPEG image
steganalysis. It needs to be fed with a great deal of labeled samples
in the training procedure. Therefore from
Fig.~\ref{fig:fig_results_train_test_all_in_boss40k_qf75}, it is no
doubt that superiority of our proposed framework in such a small-scale
dataset was not obvious. However, it still retained equal or even
slightly better performance than GFR.

\begin{table}[!t]
    \centering
    \caption[]{Comparison of number of parameters and computational complexity
      for our proposed framework and Xu's new model. The computational
      complexity is measured in terms of FLOPs~(floating-point
      operations).}
   \label{tab:xu_model_parameters}
   {\renewcommand{\arraystretch}{1.2}
     \begin{tabular}{|c|c|c|}
       \hline
       & ours with \textit{Type1} subnet & Xu's new model \\
       \hline
       Parameters & $1.66 \times 10^6$ & $4.86 \times 10^6$ \\
       \hline
       FLOPs & $2.77 \times 10^8$ & $1.53 \times 10^9$ \\
       \hline
     \end{tabular}
   }
\end{table}

\begin{figure}[!t]
  \centering
  \includegraphics[width=0.8\columnwidth,keepaspectratio]{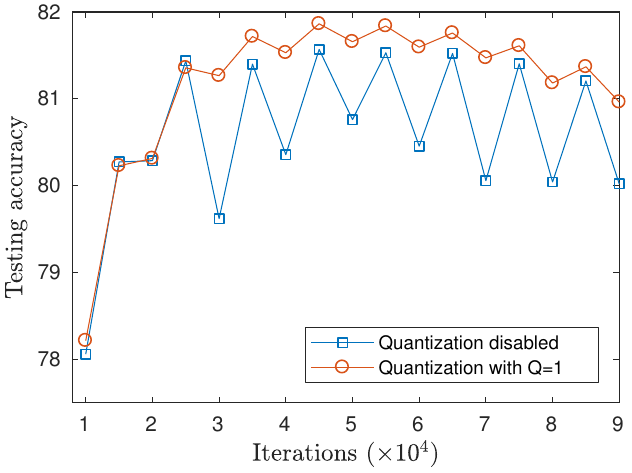}   
  \caption[]{Testing accuracies versus training iterations for Xu's
    new model~(with or without quantization $Q=1$ enabled). The
    experiment was conducted on basic500K. Only J-UNIWARD stego images
    with 0.4bpnzAC were included in the experiment. We adopted the
    training settings in Xu's work so that the training of the models
    were stopped after $9 \times 10^4$ iterations. Polyak averaging
    was enabled, as suggested by Xu.}
  \label{fig:xu_new_model} 
\end{figure}  

\begin{figure*}[!t]
  \centering
  \subfloat[]{
    \label{fig:with_xu_new_model_conceptual_architecture}
    \raisebox{-0.5\height}{\includegraphics[width=0.3\linewidth,keepaspectratio]{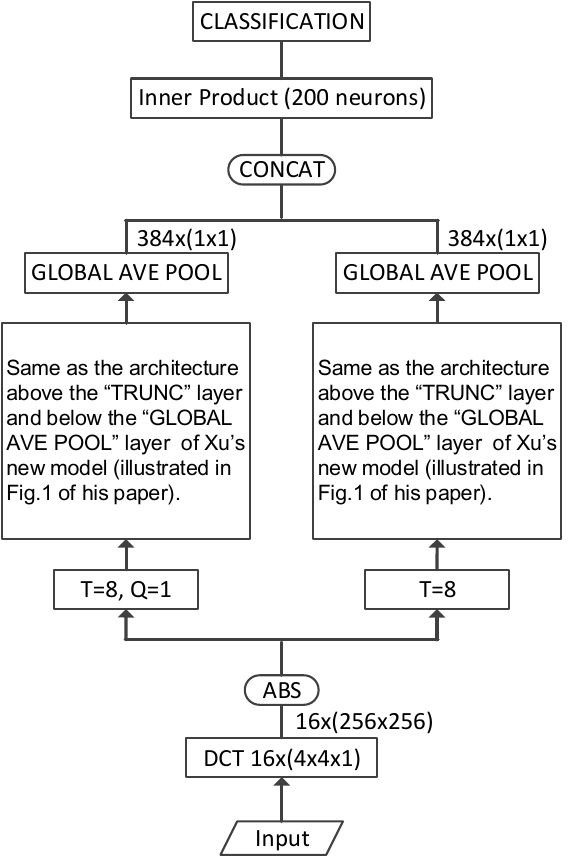}}
  }\hspace{.05\textwidth}
  \subfloat[]{
    \label{fig:xu_model_incorprated_in_our_framework}
    \raisebox{-0.5\height}{\includegraphics[width=0.4\linewidth,keepaspectratio]{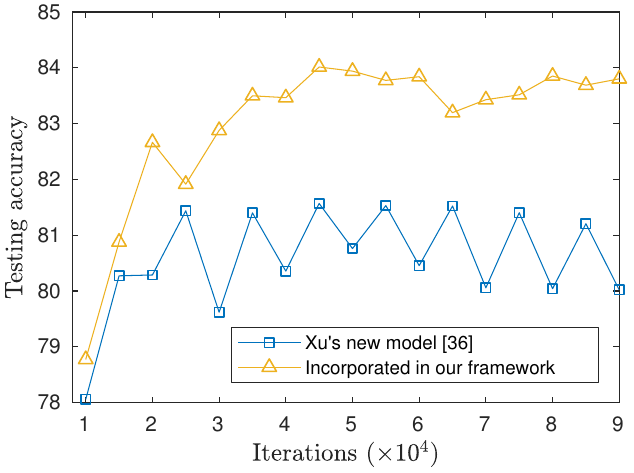}}
  }
  \caption[]{\subref{fig:with_xu_new_model_conceptual_architecture}
    Conceptual architecture of our proposed hybrid deep-learning
    framework incorporated with Xu's new model~\cite{xu_ihmmsec_2017}.
    \subref{fig:xu_model_incorprated_in_our_framework} Testing
    accuracies versus training iterations for Xu's new
    model~\cite{xu_ihmmsec_2017} and our hybrid framework incorporated
    with Xu's new model as shown in 
    \subref{fig:with_xu_new_model_conceptual_architecture}. The
    experimental setup is the same as in Fig.~\ref{fig:xu_new_model}.}
  \label{fig:our_framework_with_xu_new_model}
\end{figure*} 
\subsection{Comparison to newly emerging works}
\label{sec:exp_newly_emerging}
During the course of the review process, we noticed that two new
research works in the field of deep-learning JPEG steganalysis have
been published~\cite{xu_ihmmsec_2017,chen_ihmmsec_2017}. Due to the
limited computational capacity, we only conducted a comparative study
of our proposed framework and the framework proposed in
\cite{xu_ihmmsec_2017}~(referred as Xu's new model) since it also aims
at large-scale JPEG image steganalysis.

In \cite{xu_ihmmsec_2017}, Xu compared his framework with this work
pre-printed on arXiv, and claimed that his framework can achieve
significant performance improvement compared to the implementation of
our generic framework illustrated in Fig.~\ref{fig:1}. However, Please
note that as shown in Tab.~\ref{tab:xu_model_parameters}, Xu's new
model~\cite{xu_ihmmsec_2017} is a behemoth with about triple
parameters and more than five times of computational complexity
compared with our proposed framework with \textit{Type1}
subnet. Therefore it is natural for Xu's new
model~\cite{xu_ihmmsec_2017} to achieve better detection performance
with multiple expansions in capacity.

Xu deprecated the use of quantization in deep-learning based
steganalyzer, which we do not agree with. We conducted a verification
experiment. As shown in Fig.~\ref{fig:xu_new_model}, on the standalone
basic500K testing set which contains $500,000$ cover-stego pairs,
simply adding back quantization with $Q=1$ to Xu's new
model~\cite{xu_ihmmsec_2017} could not only make the detection
performance more stable but also improve testing accuracy.  That is to
say, even with Xu's new model~\cite{xu_ihmmsec_2017}, experimental
evidence also supports the introduction of Q\&T phase in deep-learning
steganalyzers, and supports our opinion that recognizing threshold
quantizers as a whole.

As mentioned in Sect.~\ref{sec:proposed_framework}, what proposed in
this paper is a generic hybrid architecture for deep-learning JPEG
steganalyzers. It is composed of two stages. The first stage is
equipped with hand-crafted model parameters, while the second stage is
a compound deep CNN network with a sequence of independent subnets,
and the actual number of the subnets is determined by the Q\&T
combinations experimentally. Newly emerging deep-learning
steganalyzers can be used as the prototypes of the subnets in the
second stage of our proposed hybrid architecture. Via this way, we can
incorporate them into our framework. For example, \textit{Type1}
subnet used in our work is inspired by Xu's
model~\cite{xu_spl_2016}. Certainly we can also incorporate Xu's new
model~\cite{xu_ihmmsec_2017} into our framework.  However, a complete
incorporation of Xu's new model~\cite{xu_ihmmsec_2017} in our
framework involves a great deal of experiments for architecture
adjustments~(e.g. evaluating different Q\&T combinations), and is
beyond the scope of this work. Here we just provide a straightforward
incorporation to demonstrate the generality and potentiality of our
proposed framework. As shown in
Fig.~\ref{fig:our_framework_with_xu_new_model}\subref{fig:with_xu_new_model_conceptual_architecture},
Xu's new model~\cite{xu_ihmmsec_2017} is incorporated in our hybrid
framework as the prototype of two subnets, one is with $(T=8, Q=1)$
while another is with $(T=8)$ and quantization disabled~(the original
setting in Xu's new model~\cite{xu_ihmmsec_2017}).
Fig.~\ref{fig:our_framework_with_xu_new_model}\subref{fig:xu_model_incorprated_in_our_framework}
shows the testing accuracy in successive training iterations for this
new hybrid framework and Xu's new model~\cite{xu_ihmmsec_2017}. From
Fig.~\ref{fig:our_framework_with_xu_new_model}\subref{fig:xu_model_incorprated_in_our_framework},
we can see our proposed framework incorporated with Xu's new model
outperformed the original one by a clear margin. We expect that
greater performance improvement can be achieved with more complete
incorporation of Xu's new model~\cite{xu_ihmmsec_2017} in our hybrid
generic framework.

\section{Concluding remarks}
\label{sec:concluding}

Application of deep-learning frameworks in image steganalysis has
drawn attention of many researchers. In this paper we proposed a
hybrid deep-learning framework for large-scale JPEG steganalysis,
which for the first time utilize quantization and truncation into
deep-learning steganalyzers. We have provided experimental and
theoretical testimonies to support the utilization of quantization and
truncation in the proposed framework. Our proposed framework is
generic, so that existing deep-learning based steganalyzers is easy to
be incorporated into it as a subnet prototype. We have demonstrated
the capacity of the proposed framework with different subnet
configurations, including one that incorporated from a new JPEG
deep-learning steganalyzer emerged during the review process.  The
extensive experiments conducted on a large-scale dataset extracted
from ImageNet clearly show that our proposed framework provides a
boost of performances with quantitative metrics.

Our future work will focus on two aspects: (1) incorporation of
adversarial machine learning into our proposed framework to make it
jointly optimized with its opponent; (2) further exploration of the
application of our proposed framework in the field of multimedia
forensics. 

\appendices

\section{Theoretical reflection}
\label{app:reflection}


State-of-the-art steganalytic feature extractors, either in spatial
domain or in JPEG domain, take the spatial representation~(usually
type-casted to real) of target image as
input~\cite{fridrich_tifs_2012,
  denemark_wifs_2014,tang_tifs_2016,holub_tifs_2014,
  holub_spie_2015,song_ihmmsec_2015,denemark_tifs_2016,tan_apsipa_2014,
  qian_spie_2015,qian_icip_2016,pibre_ei_2016_ver2,xu_spl_2016,
  xu_ihmmsec_2016,sedighi_ei_2017}.  Furthermore, please note that
JPEG steganalytic feature extractors are usually fed with
decompressed~(non-rounded and non-truncated) JPEG images. We follow
this approach in our research. Therefore, a grayscale input image can
be represented as
$\textrm{\textbf{X}}=(x_{pq})^{M \times
  N}=\textrm{\textbf{C}}+\textrm{\textbf{N}}$,
where $\textrm{\textbf{C}}=(c_{pq})^{M \times N}$,
$c_{pq} \in \mathbb{R}$ denotes the corresponding cover image, and
$\textrm{\textbf{N}}=(n_{pq})^{M \times N}$, $n_{pq} \in \mathbb{R}$
denotes the additive stego noise~\footnote{For JPEG steganography, the
  additive stego noise is directly added to quantized DCT
  coefficients. However, the linearity property of the DCT/IDCT
  transform guarantees that the corresponding stego noise in the
  spatial-domain representation is still additive.}.

Our reflection starts from one easily-verified fact: the magnitude
of most of the elements of $\textrm{\textbf{N}}$ matrix remain tiny
with respect to the corresponding elements of $\textrm{\textbf{C}}$
even for a stego image with high embedding rate~(on average close to
two orders of magnitude larger). State-of-the-art content-adaptive
steganography, whether in spatial domain or in JPEG domain, tends to
embed secret bits in highly textured area. As a result, even filtered
by state-of-the-art steganalytic kernels~(e.g. KV kernel used in
\cite{qian_spie_2015,pibre_ei_2016_ver2,xu_spl_2016}) the magnitudes
of most of the filtered residual elements are still much larger than
the corresponding stego noises.

Suppose that we apply a convolutional layer with kernels of size of
$m \times n$, and suppose we take as input $\textrm{\textbf{X}}$ of
size $M \times N$. Since in the context the input and output of a
convolutional layer are of two-dimensions, we adopt two-dimensional
indexing here. Convolution is just a dot product with
local-connected-and-shared weights. That is to say, for each given
$z_{rs}^{(2)}$, it is only the weighted sum of lower-layer inputs
located in a $m \times n$ local area with index $(r,s)$ as its centre
irrespective of boundary condition, and the weights used in the
weighted sum are shared in the calculation of all the $z_{rs}^{(2)}$,
$1 \le r \le M$, $1 \le s \le N$.

By rewriting \eqref{eq:2.1-3} using two-dimensional indexing, setting
$l=1$, $a_{pq}^{(1)}=x_{pq}$ and restrict the size of the dot product
to $m \times n$~($m$ and $n$ assume to be odd to omit unimportant
details), we get:
\begin{IEEEeqnarray}{rl}
  \label{eq:2.1-7}
  & z_{rs}^{(2)} =
  \sum_{p=1}^{M}\sum_{q=1}^{N}W_{pq,rs}^{(1)}x_{pq}+b_{rs}^{(1)}
  \IEEEnonumber\\
  = & \sum_{p=1}^{m}\sum_{q=1}^{n}W_{(r-\lceil \frac{m}{2}
    \rceil+p)(s-\lceil \frac{n}{2} \rceil+q),rs}^{(1)}c_{(r-\lceil
    \frac{m}{2} \rceil+p)(s-\lceil \frac{n}{2} \rceil+q)}+
  \IEEEnonumber\\
  & \ \sum_{p=1}^{m}\sum_{q=1}^{n}W_{(r-\lceil \frac{m}{2}
    \rceil+p)(s-\lceil \frac{n}{2} \rceil+q),rs}^{(1)}n_{(r-\lceil
    \frac{m}{2} \rceil+p)(s-\lceil \frac{n}{2}
    \rceil+q)}+b_{rs}^{(1)}
\end{IEEEeqnarray} 
In \eqref{eq:2.1-7}, $\lceil \cdot \rceil$ denotes the ceiling
operation. From \eqref{eq:2.1-7} we can see that if the convolutional
layer is initialized with kernels which are already sensitive to the
stego noise~(e.g. KV kernel) or is regularized as high-pass as
proposed in \cite{bayar_ihmmsec_2016}, then
$\sum_{p=1}^{m}\sum_{q=1}^{n}W_{(r-\lceil \frac{m}{2}
  \rceil+p)(s-\lceil \frac{n}{2} \rceil+q),rs}^{(1)}c_{(r-\lceil
  \frac{m}{2} \rceil+p)(s-\lceil \frac{n}{2} \rceil+q)}$
can be suppressed. However, as we mentioned above, the magnitudes of
most of the filtered residual elements are still much larger than the
corresponding stego noises, and the accumulation in \eqref{eq:2.1-7}
helps reduce the influence of outliers. Therefore in either scenario,
on average
$\sum_{p=1}^{m}\sum_{q=1}^{n}W_{(r-\lceil \frac{m}{2}
  \rceil+p)(s-\lceil \frac{n}{2} \rceil+q),rs}^{(1)}c_{(r-\lceil
  \frac{m}{2} \rceil+p)(s-\lceil \frac{n}{2} \rceil+q)}$
still accounts for the vast majority magnitude when compared with
$\sum_{p=1}^{m}\sum_{q=1}^{n}W_{(r-\lceil \frac{m}{2}
  \rceil+p)(s-\lceil \frac{n}{2} \rceil+q),rs}^{(1)}n_{(r-\lceil
  \frac{m}{2} \rceil+p)(s-\lceil \frac{n}{2} \rceil+q)}$
in \eqref{eq:2.1-7}.

For a given index $(\hat{p},\hat{q})$ where
$\hat{p}=r-\lceil \frac{m}{2} \rceil+p$,
$\hat{q}=s-\lceil \frac{n}{2} \rceil+q$, according to
\eqref{eq:2.1-vartheta_j} and \eqref{eq:2.1-5} we can see that when
the gradient is backpropagated to the layer $L_1$:
\begin{equation}
  \label{eq:gradient-l1}
  \frac{\partial}{\partial
    W_{\hat{p}\hat{q},rs}^{(1)}}J(W,b;x^{(h)},y^{(h)})=x_{\hat{p}\hat{q}}\cdot \vartheta_{rs}^{(2)}=(c_{\hat{p}\hat{q}}+n_{\hat{p}\hat{q}})\cdot \vartheta_{rs}^{(2)}
\end{equation}
in which:
\begin{equation}
  \label{eq:partial_derivative_z2}
  \vartheta_{rs}^{(2)}=(\sum_{k}W_{rs,k}^{(2)}\vartheta_{k}^{(3)})f'(z_{rs}^{(2)})
\end{equation}
In \eqref{eq:partial_derivative_z2}
$\sum_{k}W_{rs,k}^{(2)}\vartheta_{k}^{(3)}$ is fixed when the gradient
is backpropagated to the layer $L_2$. As a result
$\vartheta_{rs}^{(2)} \propto f'(z_{rs}^{(2)})$.  Please note that
$f'(z_{rs}^{(2)})$ is the derivative of the activation function of
$z_{rs}^{(2)}$. The derivatives of all of the existing practical
activation functions, including Sigmoid, TanH, and ReLU, have narrow
ranges. And furthermore, if only consider the curve in positive
axis~(or negative axis), it is easy to verify that they are linear, or
quasi-linear, namely:
\begin{equation}
  \label{eq:quasi-linear}
  \min\{f'(z_1),f'(z_2)\} \le f'(\lambda z_1+(1-\lambda)z_2) \le
  \max\{f'(z_1),f'(z_2)\},    
\end{equation}
for any $\lambda \in (0,1)$ and $z_1 \ne z_2$. Based on the fact that
$\vartheta_{rs}^{(2)} \propto f'(z_{rs}^{(2)})$,
$\vartheta_{rs}^{(2)}$ is proportional/inverse proportional to, or
quasi-proportional/inverse quasi-proportional to $z_{rs}^{(2)}$
provided the polarity of $z_{rs}^{(2)}$ remains the same. Return to
\eqref{eq:2.1-7}. Since
$\sum_{p=1}^{m}\sum_{q=1}^{n}W_{(r-\lceil \frac{m}{2}
  \rceil+p)(s-\lceil \frac{n}{2} \rceil+q),rs}^{(1)}c_{(r-\lceil
  \frac{m}{2} \rceil+p)(s-\lceil \frac{n}{2} \rceil+q)}$
accounts for the vast majority magnitude, with or without
$\sum_{p=1}^{m}\sum_{q=1}^{n}W_{(r-\lceil \frac{m}{2}
  \rceil+p)(s-\lceil \frac{n}{2} \rceil+q),rs}^{(1)}n_{(r-\lceil
  \frac{m}{2} \rceil+p)(s-\lceil \frac{n}{2} \rceil+q)}$,
the polarity of $z_{rs}^{(2)}$ will not change. Therefore the
linearity~(quasi-linearity) between $\vartheta_{rs}^{(2)}$ and
$z_{rs}^{(2)}$ holds. Consequently, due to the
linearity~(quasi-linearity) between $\vartheta_{rs}^{(2)}$ and
$z_{rs}^{(2)}$, the magnitude of $\vartheta_{rs}^{(2)}$ mainly depends
on the weighted sum of the cover image pixels located in the
corresponding $m \times n$ local area, rather than the weighted sum of
those stego noises.

Furthermore, in \eqref{eq:gradient-l1} we can see there is a multiply
factor to $\vartheta_{rs}^{(2)}$,
$(c_{\hat{p}\hat{q}}+n_{\hat{p}\hat{q}})$. Since by average
$|c_{\hat{p}\hat{q}}|$ is close to two orders of magnitude larger than
$|n_{\hat{p}\hat{q}}|$ even with a high embedding rate, the impact of
the neighboring cover image pixels on
$\frac{\partial}{\partial
  W_{\hat{p}\hat{q},rs}^{(1)}}J(W,b;x^{(h)},y^{(h)})$
is further amplified. As a result, the influence of
$n_{\hat{p}\hat{q}}$, and the neighboring stego noise in the
corresponding $m \times n$ local area, to
$\frac{\partial}{\partial
  W_{\hat{p}\hat{q},rs}^{(1)}}J(W,b;x^{(h)},y^{(h)})$
becomes very weak.  At last, since in a convolutional layer the
weights are shared, all the partial derivatives with respect to a
given shared weight should be accumulated:
\begin{multline}
  \label{eq:kernel_accumulated}
  \frac{\partial}{\partial
    W_{pq}^{(1)}}J(W,b;x^{(h)},y^{(h)})=
  \sum_{r=1}^{M}\sum_{s=1}^{N} \frac{\partial
    J(W,b;x^{(h)},y^{(h)})}{\partial W_{(r-\lceil \frac{m}{2}
      \rceil+p)(s-\lceil \frac{n}{2}
      \rceil+q),rs}^{(1)}},\\
  1 \le p \le m,\ 1 \le q \le n.
\end{multline}
The accumulation in \eqref{eq:kernel_accumulated} again helps reduce
the influence of outliers. As a result, it is safe for us to make a
conclusion that the influence of stego noises to
$\frac{\partial}{\partial W_{pq}^{(1)}}J(W,b;x^{(h)},y^{(h)})$,
$1 \le p \le m,\ 1 \le q \le n$ is weak in statistical sense.
Consequently, gradient descent algorithm in the bottom convolutional
layer will be always guided by the cover image contents rather than
the stego noises. In other words, the optimization of the bottom
convolutional layer in favor of the extraction of stego noises is hard
to achieve with gradient descent.

\section*{Acknowledgment}

The authors would like to thank DDE Laboratory in SUNY Binghamton and
Dr.~Guanshuo Xu for sharing the source code of their steganalysis
models online. We also appreciate Prof. Jiangqun Ni in Sun Yat-sen
University, China for permission to use their implementation of UED
and UERD in our experiments. Specifically, we are grateful to Dr.
Pawe\l\ Korus at that time in Shenzhen University for valuable advice.


\bibliographystyle{IEEEtran}
\bibliography{IEEEfull,jsdl_final}

\begin{thebibliography}{10}
\providecommand{\url}[1]{#1}
\csname url@samestyle\endcsname
\providecommand{\newblock}{\relax}
\providecommand{\bibinfo}[2]{#2}
\providecommand{\BIBentrySTDinterwordspacing}{\spaceskip=0pt\relax}
\providecommand{\BIBentryALTinterwordstretchfactor}{4}
\providecommand{\BIBentryALTinterwordspacing}{\spaceskip=\fontdimen2\font plus
\BIBentryALTinterwordstretchfactor\fontdimen3\font minus
  \fontdimen4\font\relax}
\providecommand{\BIBforeignlanguage}[2]{{%
\expandafter\ifx\csname l@#1\endcsname\relax
\typeout{** WARNING: IEEEtran.bst: No hyphenation pattern has been}%
\typeout{** loaded for the language `#1'. Using the pattern for}%
\typeout{** the default language instead.}%
\else
\language=\csname l@#1\endcsname
\fi
#2}}
\providecommand{\BIBdecl}{\relax}
\BIBdecl

\bibitem{fridrich_spie_2007}
J.~Fridrich and T.~Filler, ``Practical methods for minimizing embedding impact
  in steganography,'' in \emph{Proc. SPIE, Electronic Imaging, Security,
  Steganography, and Watermarking of Multimedia Contents IX}, vol. 6505, 2007,
  pp. 650\,502--1--650\,502--15.

\bibitem{pevny_ih_2010}
T.~Pevn\'y, T.~Filler, and P.~Bas, ``Using high-dimensional image models to
  perform highly undetectable steganography,'' in \emph{Proc. 12th Information
  Hiding Workshop ({IH}'2010)}, 2010, pp. 161--177.

\bibitem{li_icip_2014}
B.~Li, M.~Wang, J.~Huang, and X.~Li, ``A new cost function for spatial image
  steganography,'' in \emph{Proc. IEEE 2014 International Conference on Image
  Processing, ({ICIP}'2014)}, 2014, pp. 4206--4210.

\bibitem{sedighi_tifs_2016}
V.~Sedighi, R.~Cogranne, and J.~Fridrich, ``Content-adaptive steganography by
  minimizing statistical detectability,'' \emph{{IEEE} Transactions on
  Information Forensics and Security}, vol.~11, no.~2, pp. 221--234, 2016.

\bibitem{guo_tifs_2014}
L.~Guo, J.~Ni, and Y.~Q. Shi, ``Uniform embedding for efficient {JPEG}
  steganography,'' \emph{{IEEE} Transactions on Information Forensics and
  Security}, vol.~9, no.~5, pp. 814--825, 2014.

\bibitem{guo_tifs_2015}
L.~Guo, J.~Ni, W.~Su, C.~Tang, and Y.~Q. Shi, ``Using statistical image model
  for {JPEG} steganography: Uniform embedding revisited,'' \emph{{IEEE}
  Transactions on Information Forensics and Security}, vol.~10, no.~12, pp.
  2669--2680, 2015.

\bibitem{holub_eurasip_2014}
V.~Holub, J.~Fridrich, and T.~Denemark, ``Universal distortion function for
  steganography in an arbitrary domain,'' \emph{EURASIP Journal on Information
  Security}, vol. 2014, no.~1, pp. 1--13, 2014.

\bibitem{denemark_ihmmsec_2015}
T.~Denemark and J.~Fridrich, ``Improving steganographic security by
  synchronizing the selection channel,'' in \emph{Proc. 3rd ACM Information
  Hiding and Multimedia Security Workshop (IH\&MMSec'2015)}, 2015, pp. 5--14.

\bibitem{li_tifs_2015}
B.~Li, M.~Wang, X.~Li, S.~Tan, and J.~Huang, ``A strategy of clustering
  modification directions in spatial image steganography,'' \emph{{IEEE}
  Transactions on Information Forensics and Security}, vol.~10, no.~9, pp.
  1905--1917, 2015.

\bibitem{denemark_wifs_2015}
T.~Denemark and J.~Fridrich, ``Side-informed steganography with additive
  distortion,'' in \emph{Proc. 7th IEEE International Workshop on Information
  Forensic and Security ({WIFS}'2015)}, 2015, pp. 1--6.

\bibitem{fridrich_tifs_2012}
J.~Fridrich and J.~Kodovsk\'y, ``Rich models for steganalysis of digital
  images,'' \emph{{IEEE} Transactions on Information Forensics and Security},
  vol.~7, no.~3, pp. 868--882, 2012.

\bibitem{denemark_wifs_2014}
T.~Denemark, V.~Sedighi, V.~Holub, R.~Cogranne, and J.~Fridrich,
  ``Selection-channel-aware rich model for steganalysis of digital images,'' in
  \emph{Proc. 6th IEEE International Workshop on Information Forensic and
  Security ({WIFS}'2014)}, 2014, pp. 48--53.

\bibitem{tang_tifs_2016}
W.~Tang, H.~Li, W.~Luo, and J.~Huang, ``Adaptive steganalysis based on
  embedding probabilities of pixels,'' \emph{{IEEE} Transactions on Information
  Forensics and Security}, vol.~11, no.~4, pp. 734--745, 2016.

\bibitem{kodovsky_tifs_2012}
J.~Kodovsk\'y and J.~Fridrich, ``Ensemble classifiers for steganalysis of
  digital media,'' \emph{{IEEE} Transactions on Information Forensics and
  Security}, vol.~7, no.~2, pp. 432--444, 2012.

\bibitem{holub_tifs_2014}
V.~Holub and J.~Fridrich, ``Low-complexity features for {JPEG} steganalysis
  using undecimated {DCT},'' \emph{{IEEE} Transactions on Information Forensics
  and Security}, vol.~10, no.~2, pp. 219--228, 2015.

\bibitem{holub_spie_2015}
------, ``Phase-aware projection model for steganalysis of {JPEG} images,'' in
  \emph{Proc. IS\&T/SPIE Electronic Imaging 2015~(Media Watermarking, Security,
  and Forensics)}, 2015, pp. 94\,090T--1--94\,090T--11.

\bibitem{song_ihmmsec_2015}
X.~Song, F.~Liu, C.~Yang, X.~Luo, and Y.~Zhang, ``Steganalysis of adaptive
  {JPEG} steganography using 2d {Gabor} filters,'' in \emph{Proc. 3rd ACM
  Information Hiding and Multimedia Security Workshop (IH\&MMSec'2015)}, 2015,
  pp. 15--23.

\bibitem{denemark_tifs_2016}
T.~Denemark, M.~Boroumand, and J.~Fridrich, ``Steganalysis features for
  content-adaptive {JPEG} steganography,'' \emph{{IEEE} Transactions on
  Information Forensics and Security}, vol.~11, no.~8, pp. 1736--1746, 2016.

\bibitem{schmidhuber_nn_2015}
J.~Schmidhuber, ``Deep learning in neural networks: An overview,'' \emph{Neural
  Networks}, vol.~61, pp. 85--117, 2015.

\bibitem{tan_apsipa_2014}
S.~Tan and B.~Li, ``Stacked convolutional auto-encoders for steganalysis of
  digital images,'' in \emph{Proc. Asia-Pacific Signal and Information
  Processing Association Annual Summit and Conference ({APSIPA}'2014)}, 2014.

\bibitem{qian_spie_2015}
Y.~Qian, J.~Dong, W.~Wang, and T.~Tan, ``Deep learning for steganalysis via
  convolutional neural networks,'' in \emph{Proc. IS\&T/SPIE Electronic Imaging
  2015~(Media Watermarking, Security, and Forensics)}, 2015, pp.
  94\,090J--1--94\,090J--10.

\bibitem{qian_icip_2016}
------, ``Learning and transferring representations for image steganalysis
  using convolutional neural network,'' in \emph{Proc. IEEE 2016 International
  Conference on Image Processing, ({ICIP}'2016)}, 2016, pp. 2752--2756.

\bibitem{pibre_ei_2016_ver2}
L.~Pibre, P.~J\'{e}r\^{o}me, D.~Ienco, and M.~Chaumont, ``Deep learning is a
  good steganalysis tool when embedding key is reused for different images,
  even if there is a cover source-mismatch,'' in \emph{Proc. Media
  Watermarking, Security, and Forensics, Part of {IS\&T} International
  Symposium on Electronic Imaging (EI'2016)}, San Francisco, CA, USA, 14-18
  February 2016.

\bibitem{xu_spl_2016}
G.~Xu, H.~Z. Wu, and Y.~Q. Shi, ``Structural design of convolutional neural
  networks for steganalysis,'' \emph{{IEEE} Signal Processing Letters},
  vol.~23, no.~5, pp. 708--712, 2016.

\bibitem{xu_ihmmsec_2016}
------, ``Ensemble of {CNNs} for steganalysis: {An} empirical study,'' in
  \emph{Proc. 4th ACM Information Hiding and Multimedia Security Workshop
  (IH\&MMSec'2016)}, 2016, pp. 103--107.

\bibitem{ioffe_arxiv_2015}
\BIBentryALTinterwordspacing
S.~Ioffe and C.~Szegedy, ``Batch normalization: Accelerating deep network
  training by reducing internal covariate shift,'' \emph{arXiv:1502.03167},
  2015. [Online]. Available: \url{http://arxiv.org/abs/1502.03167}
\BIBentrySTDinterwordspacing

\bibitem{sedighi_ei_2017}
V.~Sedighi and J.~Fridrich, ``Histogram layer, moving convolutional neural
  networks towards feature-based steganalysis,'' in \emph{Proc. Media
  Watermarking, Security, and Forensics, Part of {IS\&T} International
  Symposium on Electronic Imaging (EI'2017)}, Burlingame, CA, 29 Juanuary- 2
  February 2017.

\bibitem{bas_ih_2011}
P.~B.~T. Filler and T.~Pevn\'y, ``Break our steganographic system---the ins and
  outs of organizing {BOSS},'' in \emph{Proc. 13th Information Hiding Workshop
  ({IH}'2011)}, 2011, pp. 59--70.

\bibitem{sedighi_spie_2016}
V.~Sedighi, J.~Fridrich, and R.~Cogranne, ``Toss that {BOSSbase}, {Alice}!'' in
  \emph{Proc. Media Watermarking, Security, and Forensics, Part of {IS\&T}
  International Symposium on Electronic Imaging (EI'2016)}, San Francisco, CA,
  USA, 14-18 February 2016.

\bibitem{zeng_ei_2017}
J.~Zeng, S.~Tan, and B.~Li, ``Pre-training via fitting deep neural network to
  rich-model features extraction procedure and its effect on deep learning for
  steganalysis,'' in \emph{Proc. Media Watermarking, Security, and Forensics,
  Part of {IS\&T} International Symposium on Electronic Imaging (EI'2017)},
  Burlingame, CA, USA, 29 Juanuary- 2 February 2017.

\bibitem{imagenet_website}
``{ImageNet},'' \url{http://image-net.org/}, [Online].

\bibitem{cs231n_website}
``{CS231n: Convolutional Neural Networks for Visual Recognition},''
  \url{http://cs231n.github.io/}, [Online].

\bibitem{szegedy_cvpr_2015}
C.~Szegedy, W.~Liu, Y.~Jia, P.~Sermanet, S.~Reed, D.~Anguelov, D.~Erhan,
  V.~Vanhoucke, and A.~Rabinovich, ``Going deeper with convolutions,'' in
  \emph{Proc. IEEE Conference on Computer Vision and Pattern Recognition},
  2015, pp. 1--9.

\bibitem{bayar_ihmmsec_2016}
B.~Bayar and M.~C. Stamm, ``A deep learning approach to universal image
  manipulation detection using a new convolutional layer,'' in \emph{Proc. 4th
  ACM Information Hiding and Multimedia Security Workshop (IH\&MMSec'2016)},
  2016, pp. 5--10.

\bibitem{jia_arxiv_2014}
\BIBentryALTinterwordspacing
Y.~Jia, E.~Shelhamer, J.~Donahue, S.~Karayev, J.~Long, R.~Girshick,
  S.~Guadarrama, and T.~Darrell, ``Caffe: Convolutional architecture for fast
  feature embedding,'' \emph{arXiv:1408.5093}, 2014. [Online]. Available:
  \url{http://arxiv.org/abs/1408.5093}
\BIBentrySTDinterwordspacing

\bibitem{xu_ihmmsec_2017}
G.~Xu, ``Deep convolutional neural network to detect {J-UNIWARD},'' in
  \emph{Proc. 5th ACM Information Hiding and Multimedia Security Workshop
  (IH\&MMSec'2017)}, 2017, pp. 67--73.

\bibitem{chen_ihmmsec_2017}
M.~Chen, V.~Sedighi, M.~Boroumand, and J.~Fridrich, ``{JPEG}-phase-aware
  convolutional neural network for steganalysis of {JPEG} images,'' in
  \emph{Proc. 5th ACM Information Hiding and Multimedia Security Workshop
  (IH\&MMSec'2017)}, 2017, pp. 75--84.

\end{thebibliography}

\end{document}